\documentclass[letterpaper]{article} 
\usepackage{aaai2027}
\usepackage[hyphens]{url}  
\usepackage{graphicx} 
\usepackage{natbib}  
\usepackage{caption} 
\usepackage{algorithm}
\usepackage{algorithmic}
\usepackage{multirow}
\usepackage{booktabs}
\usepackage{pifont}
\usepackage{amsmath}
\usepackage{bm}
\usepackage{enumitem}
\usepackage{amssymb}
\usepackage{amsmath}
\usepackage{amsthm}

\usepackage{makecell}

\usepackage{xcolor}
\usepackage[most]{tcolorbox}
\newtcolorbox{promptbox}[1]{
  colback=white,
  colframe=gray,
  coltitle=white,
  fonttitle=\bfseries,
  title=#1,
  arc=4pt,
  boxrule=1.2pt,
  left=8pt,
  right=8pt,
  top=6pt,
  bottom=6pt
}
\newtcolorbox{casebox}[1]{
  enhanced,
  breakable,
  colback=white,
  colframe=gray,
  coltitle=white,
  fonttitle=\bfseries,
  title=#1,
  arc=4pt,
  boxrule=1.2pt,
  left=8pt,
  right=8pt,
  top=6pt,
  bottom=6pt,
  fontupper=\small
}

\usepackage{newfloat}
\usepackage{listings}
\DeclareCaptionStyle{ruled}{labelfont=normalfont,labelsep=colon,strut=off} 
\floatstyle{ruled}
\newfloat{listing}{tb}{lst}{}
\floatname{listing}{Listing}

\usepackage{booktabs}

\nocopyright

\title{When Semantics Fall Short: Persona-Guided Cross-Domain Sequential Recommendation with Large Language Models}
\author{
    Jundong Chen\textsuperscript{\rm 1,2},
    Honglei Zhang\textsuperscript{\rm 1,2}\thanks{Corresponding author.},
    Shuai An\textsuperscript{\rm 1,2},
    Enneng Yang\textsuperscript{\rm 3},
    Han Yu\textsuperscript{\rm 4}
}
\affiliations{
    \textsuperscript{\rm 1}Key Laboratory of Big Data \& Artificial Intelligence in Transportation, Ministry of Education, China\\
    \textsuperscript{\rm 2}Beijing Jiaotong University\quad\textsuperscript{\rm 3}Sun Yat-sen University\quad\textsuperscript{\rm 4}Nanyang Technological University\\
    \{jundongchen, honglei.zhang, shuaian\}@bjtu.edu.cn, ennengyang@stumail.neu.edu.cn, han.yu@ntu.edu.sg
}

\begin{document}

\maketitle

\begin{abstract}
Cross-domain sequential recommendation (CDSR) alleviates data sparsity by transferring user behaviors across domains. Recent advances increasingly leverage large language models (LLMs) to bridge domains through global item semantics. However, this paradigm is inherently limited by two issues: I. semantic domain isolation, where item semantics remain domain-specific and hinder transferable modeling, and II. semantic-behavior inconsistency, where local semantic relations are more consistent with behavioral transitions, whereas such consistency often breaks down globally across domains. To this end, we propose PerCDR, an LLM-enhanced CDSR framework that shifts the focus from global semantic bridging to persona-guided behavioral modeling. Specifically, PerCDR leverages LLM-inferred personas to globally guide transferable cross-domain behavioral patterns, while exploiting local semantic relations to enhance intra-domain behavioral modeling, followed by domain preference adaptation for recommendation refinement. Extensive experiments demonstrate the superiority of PerCDR and validate its effectiveness. Our code is available at \url{https://github.com/jundongchen13/PerCDR}.
\end{abstract}


\section{Introduction}
Sequential Recommendation (SR) aims to predict a user’s next interaction based on historical behavior sequences and has become a fundamental paradigm for personalized recommendation~\cite{kang2018self, pan2026survey}. However, the inherent sparsity of user-item interactions often results in insufficient behavioral modeling, thereby limiting model performance. To alleviate this issue, Cross-Domain Sequential Recommendation (CDSR) exploits user behaviors from auxiliary domains to improve user modeling in the target domain, emerging as a promising solution~\cite{cao2022contrastive, park2024pacer, chen2024survey, chen2026transfer}.

\begin{figure*}[htbp]
\centering
\includegraphics[width=1.0\textwidth]{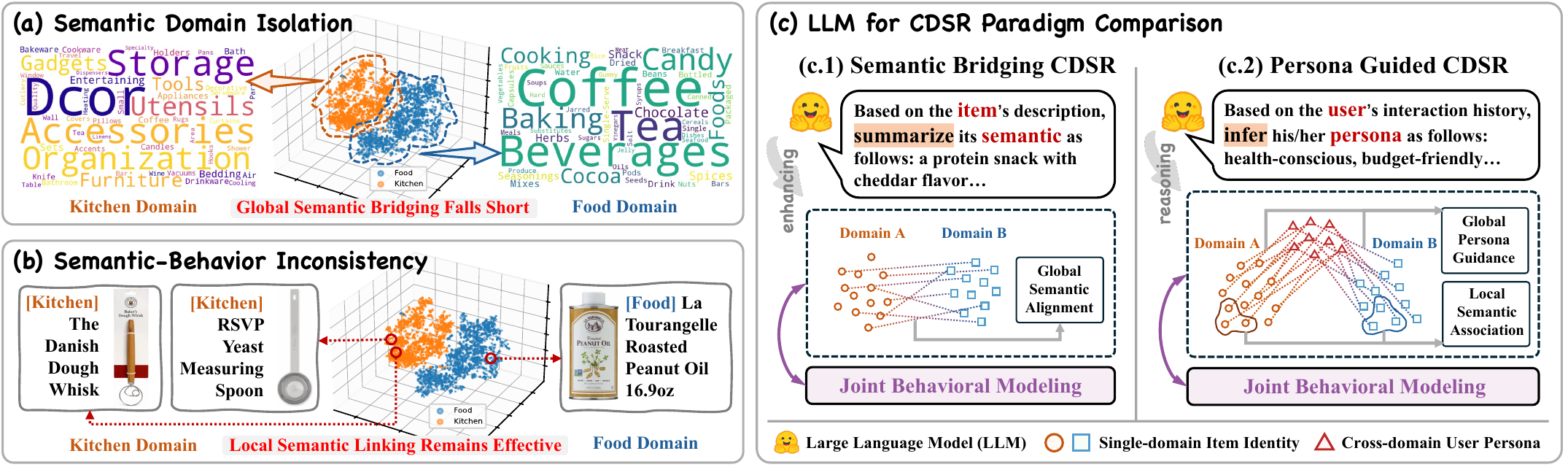}
\caption{Motivation and paradigm comparison. Existing LLM-enhanced CDSR methods adopt an item-centric perspective through global semantic bridging, which suffers from semantic domain isolation and semantic-behavior inconsistency. PerCDR instead adopts a user-centric perspective, leveraging global persona guidance for cross-domain behavioral modeling.}
\label{pic:motivation}
\end{figure*}

Classic CDSR methods primarily rely on collaborative signals from user-item interactions for cross-domain transfer. For instance, C$^2$DSR~\cite{cao2022contrastive} captures cross-domain item relations through interaction co-occurrences, and Tri-CDR~\cite{ma2024triple} enhances user modeling with mixed-domain behavioral sequences. However, such signals are often too scarce to establish reliable relations across heterogeneous domains. Consequently, some studies have incorporated textual information into CDSR, using item semantics to establish cross-domain connections. Early efforts, such as UniSRec~\cite{hou2022towards} and CTT~\cite{zhou2025contrastive}, leverage pre-trained language models to enrich item representations and uncover transferable knowledge across domains. More recently, the emergence of Large Language Models (LLMs) has further enhanced semantic learning, inspiring approaches such as LLMCDSR~\cite{xin2025llmcdsr}, LLM4CDSR~\cite{liu2025bridge}, and SemaCDR~\cite{zhang2026semacdr}. Despite their effectiveness, these methods share a common \texttt{semantic bridging} paradigm, as illustrated in Figure~\ref{pic:motivation}(c.1), which globally aligns item semantics to facilitate joint behavioral modeling across domains.

However, such a paradigm suffers from two key issues:\\
\textbf{Issue I. Semantic Domain
Isolation}. As illustrated in Figure~\ref{pic:motivation}(a), items in the Kitchen domain are primarily described by functional attributes, \textit{e.g.}, “storage” and “accessories”, whereas items in the Food domain are characterized by product categories, \textit{e.g.}, “coffee” and “tea”. Such isolated domain-specific semantics make it difficult for global alignment to learn transferable behavioral patterns across domains. \\
\textbf{Issue II. Semantic-Behavior Inconsistency}. As shown in Figure~\ref{pic:motivation}(b), a user who purchases a spoon subsequently interacts with a semantically related item within the same domain, \textit{e.g.}, a whisk, but also purchases a semantically distant item in another domain, \textit{e.g.}, peanut oil. This observation suggests that semantic proximity is more consistent with behavioral transitions within a domain than across domains.

Taken together, these observations suggest that the conventional \emph{item-centric} perspective, which relates domains through global item semantics, is inherently suboptimal. Instead, CDSR should adopt a \emph{user-centric} perspective that relates behaviors across domains according to the underlying behavioral rationale of users. Motivated by this insight, we propose a \texttt{persona-guided} CDSR paradigm, as illustrated in Figure~\ref{pic:motivation}(c.2). Specifically, we leverage LLMs to infer transferable user personas from mixed-domain interaction histories, capturing high-level user characteristics such as habits, intents, and behavioral tendencies. These user personas then provide global guidance for establishing cross-domain behavioral relations, rather than globally aligning item semantics. Meanwhile, local semantic structures are exploited as associations to strengthen intra-domain behavioral modeling. By leveraging global persona guidance and local semantic association, our framework performs joint modeling to learn transferable behavioral patterns across domains. Furthermore, we introduce a domain preference adaptation strategy that refines the joint predictions with domain-specific predictions for more accurate recommendations.
To sum up, our contributions are as follows:
\begin{itemize}
    \item We propose PerCDR, a novel persona-guided CDSR framework that moves beyond the conventional semantic bridging paradigm. It consists of two stages: Joint Behavioral Modeling and Domain Preference Adaptation.
    \item In the Joint Behavioral Modeling stage, LLM-inferred user personas provide global guidance for cross-domain behavioral modeling, while local semantic associations enhance intra-domain behavioral modeling, enabling unified behavioral patterns to be learned across domains.
    \item In the Domain Preference Adaptation stage, joint predictions are refined with domain-specific ones to better adapt to domain preferences and improve performance.
    \item Extensive experiments demonstrate the superiority of PerCDR over state-of-the-art (SOTA) methods, and further analyses validate the effectiveness of the proposed model.
\end{itemize}

\section{Related Work}
\subsection{Cross-domain Sequential Recommendation} 
Building upon Sequential Recommendation (SR), Cross-Domain Sequential Recommendation (CDSR) leverages user behaviors from auxiliary domains to alleviate interaction sparsity and improve user modeling. Early studies primarily establish cross-domain transfer through collaborative signals. Representative methods include $\pi$-Net~\cite{ma2019pi}, which learns shared user preferences via cross-domain transfer modules, C$^2$DSR~\cite{cao2022contrastive}, which models cross-domain item co-occurrences using graph neural networks, and Tri-CDR~\cite{ma2024triple}, which captures global interests from mixed-domain behavioral sequences through contrastive learning. To further facilitate cross-domain transfer, recent studies have introduced textual side information into CDSR. Representative methods such as UniSRec~\cite{hou2022towards} and CTT~\cite{zhou2025contrastive} leverage pre-trained language models to enrich item representations for transferability, while HGTL~\cite{xu2025heterogeneous} further incorporates category semantics to strengthen structural modeling. These studies demonstrate the effectiveness of textual information in CDSR and motivate the subsequent exploration of richer semantic knowledge with LLMs.

\subsection{LLM-enhanced Sequential Recommendation}
Large Language Models (LLMs) have demonstrated remarkable capabilities in semantic understanding and reasoning, driving rapid advances in SR~\cite{liu2025large, wu2024survey}. Existing studies mainly follow two paradigms. The first treats LLMs as recommenders through prompt-based inference or task-specific fine-tuning, such as TALLRec~\cite{bao2023tallrec} and TaxRec~\cite{liang2025taxonomy}. The second employs LLMs to enhance SR models by generating semantic representations or auxiliary supervision, \textit{e.g.}, SAID~\cite{hu2024enhancing}, LLMEmb~\cite{liu2025llmemb}, LLM-ESR~\cite{liu2024llm}, and CLLMRec~\cite{lu2025cllmrec}. Recent studies have further incorporated LLMs into CDSR. Representative methods either generate semantically related pseudo items, \textit{e.g.}, LLMCDSR~\cite{xin2025llmcdsr} and LLM-EDT~\cite{liu2025llm}, or enhance cross-domain semantic representations, \textit{e.g.}, LLM4CDSR~\cite{liu2025bridge} and SemaCDR~\cite{zhang2026semacdr}. Despite their differences, these methods consistently employ LLMs for item-centric semantic bridging. In contrast, we leverage LLMs to infer transferable personas, guiding user-centric cross-domain behavioral modeling.


\section{Preliminaries}
\textbf{Sequential Recommendation.} 
Given the user set $\mathcal{U}$, item set $\mathcal{I}$, and interaction sequence set $\mathcal{S}=\{s_u\}_{u\in\mathcal{U}}$, each interaction sequence is denoted by $s_u=[i_1,i_2,\cdots,i_{|s_u|}]$, where $i_t\in\mathcal{I}$ denotes the interacted item at the $t$-th timestamp. A sequential recommendation model consists of an item embedding table $\mathbf{E}\in\mathbb{R}^{|\mathcal{I}|\times d}$ and a sequential encoder $f_{\theta}(\cdot)$, where $|\mathcal{I}|$ is the number of items and $d$ is the embedding dimension. The interaction sequence is first mapped into an embedding sequence and then encoded as:
\begin{equation}\label{eq:seq_rec}
    \{\mathbf{h}_u^{(t)}\}_{t=1}^{|s_u|}=f_{\theta}(\mathbf{E}[s_u])=f_{\theta}([\mathbf{e}_1,\mathbf{e}_2,\cdots,\mathbf{e}_{|s_u|}]),
\end{equation}
where $\mathbf{e}_t=\mathbf{E}[i_t]\in\mathbb{R}^d$ denotes the embedding of item $i_t$, and $\mathbf{h}_u^{(t)}\in\mathbb{R}^d$ denotes the hidden representation at the $t$-th timestamp. Following common practice, the last hidden representation is used as the user representation, denoted by $\mathbf{h}_u=f_{\theta}(\mathbf{E}[s_u])$ for simplicity. The scores over all candidate items are computed by $\mathbf{r}_u=\mathbf{h}_u\mathbf{E}^{\top}\in\mathbb{R}^{|\mathcal{I}|}$, where the Top-$N$ items with the highest scores are recommended to the user. Throughout this paper, we instantiate $f_{\theta}(\cdot)$ using SASRec~\cite{kang2018self}.

\noindent \textbf{Cross-domain Sequential Recommendation.} We consider two domains A and B, with distinct item sets $\mathcal{I}^{A}$ and $\mathcal{I}^{B}$, respectively. The two domains share the same user set $\mathcal{U}$. The corresponding interaction sequence sets are denoted by $\mathcal{S}^{A}=\{s_u^{A}\}_{u\in\mathcal{U}}$ and $\mathcal{S}^{B}=\{s_u^{B}\}_{u\in\mathcal{U}}$. Following existing CDSR methods, the two interaction sequences are chronologically merged into a mixed-domain sequence set $\tilde{\mathcal{S}}=\{\tilde{s}_u\}_{u\in\mathcal{U}}$, where $\tilde{s}_u=[i_1^{D_1},i_2^{D_2},\cdots,i_{|\tilde{s}_u|}^{D_{|\tilde{s}_u|}}]$,
and $D_t\in\{A,B\}$ indicates the domain of the corresponding interaction. The objective of CDSR is to perform next-item prediction in both domains under a dual-sequence setting, where single-domain sequences $\mathcal{S}^{A},\mathcal{S}^{B}$ capture domain-specific preferences, while the mixed-domain sequence $\tilde{\mathcal{S}}$ models transferable cross-domain behaviors.


\section{Methodology}
\subsection{Overall Framework}
\begin{figure*}[htbp]
\centering
\includegraphics[width=1.0\textwidth]{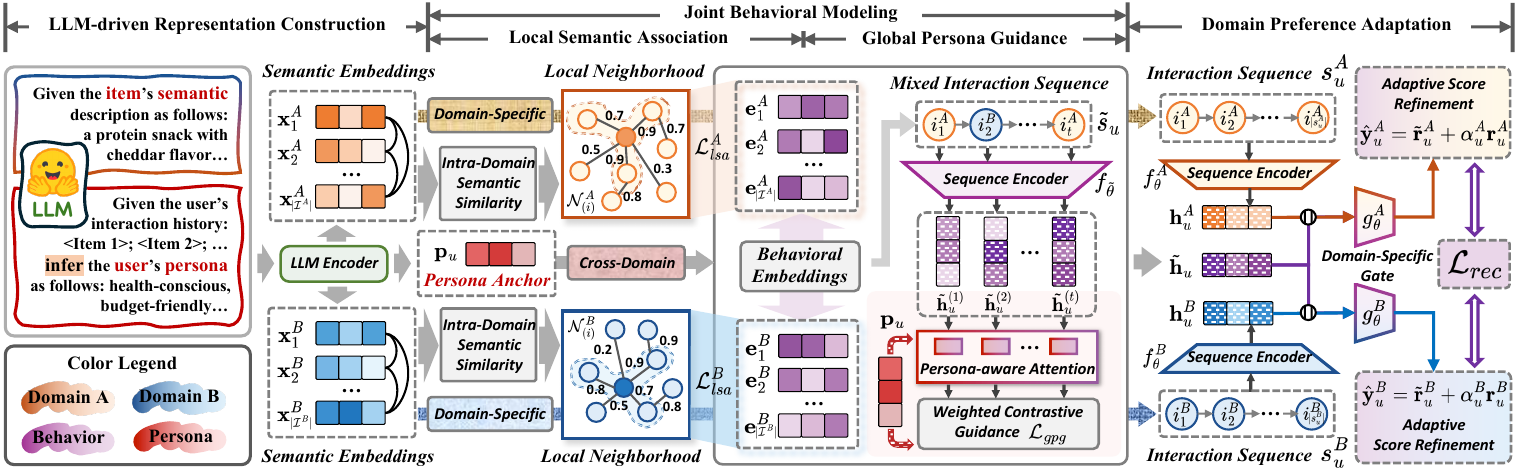}
\caption{Overall framework of PerCDR. The framework consists of three stages: (I) LLM-driven Representation Construction for semantic embedding and persona anchor generation; (II) Joint Behavioral Modeling for collaborative intra- and cross-domain behavior learning; and (III) Domain Preference Adaptation for adaptive recommendation refinement.}
\label{pic:framework}
\end{figure*}

The overall framework of PerCDR is illustrated in Figure~\ref{pic:framework}, which consists of three major stages:\\
(I) \textit{LLM-driven Representation Construction}. We first employ LLMs to infer user persona anchors and construct frozen item semantic embeddings $\mathbf{X}^{A}$ and $\mathbf{X}^{B}$.\\
(II) \textit{Joint Behavioral Modeling}. 
We jointly learn trainable behavioral embeddings $\mathbf{E}^{A}$ and $\mathbf{E}^{B}$ with a shared sequential encoder $f_{\tilde{\theta}}(\cdot)$ over the mixed-domain interaction sequence $\tilde{\mathcal S}$ to capture transferable cross-domain behaviors. During this process, frozen semantic embeddings provide local associations, while persona anchors offer global guidance.\\
(III) \textit{Domain Preference Adaptation}. Based on the single-domain interaction sequence sets $\mathcal S^{A}$ and $\mathcal S^{B}$, we employ two independent sequential encoders $f_{\theta}^{A}(\cdot)$ and $f_{\theta}^{B}(\cdot)$ over the frozen semantic embeddings $\mathbf{X}^{A}$ and $\mathbf{X}^{B}$ to capture domain-specific preferences, which are further used to adaptively refine the joint recommendation scores.

\subsection{LLM-driven Representation Construction}
\noindent \textbf{Item Semantic Identity.} We first construct semantic identities for all items using an LLM embedding model based on their textual attributes $T_{i}$, including the title, category, and description. This process can be formulated as:
\begin{equation}\label{eq:item_emb}
    \mathbf{x}_i=\mathrm{LLM}_{emb}(T_i).
\end{equation}
The resulting semantic embeddings for two domains can be denoted as $\mathbf{X}^A=\{\mathbf{x}_i^A\}_{i\in\mathcal{I}^A}$ and $\mathbf{X}^B=\{\mathbf{x}_i^B\}_{i\in\mathcal{I}^B}$.

\noindent \textbf{User Persona Anchor.} Existing CDSR methods further leverage user profiles derived from historical item semantics, which summarize \emph{what} users interacted with to enhance semantic modeling. In contrast, PerCDR leverages user personas to explain \emph{why} users interact across domains by capturing the behavioral rationale underlying user interactions. This behavioral rationale is inherently more transferable than item semantics, making it a natural prior for guiding cross-domain behavioral modeling. Specifically, we exploit an LLM reasoning model to infer the persona $P_u$ from each user’s mixed-domain interaction history $\tilde{s}_u$. The prompt template $\mathcal{F}$ is:
\begin{promptbox}{Prompt Template $\mathcal{F}$}
Given a user's chronological cross-domain interaction history: \underline{<Item Title 1>}; \underline{<Item Title 2>}; \underline{...}

Infer one transferable user persona that summarizes the user's high-level habits, intents, and decision tendencies, capturing the common behavioral rationale underlying the interactions across domains.

Output only one concise persona description. Do not mention specific items, brands, or domains.
\end{promptbox}
\noindent The persona anchor construction process is formulated as:
\begin{equation}\label{eq:user_anc}
P_u=\mathrm{LLM}_{reason}(\mathcal{F}(\tilde{s}_u)), \,\, \mathbf{p}_u=\mathrm{LLM}_{emb}(P_u).
\end{equation}
\noindent Following common practice~\cite{liu2025bridge}, both $\mathbf{x}_i$ and $\mathbf{p}_u$ are projected to the model dimension $d$ using a lightweight projection module, \textit{e.g.}, a Multi-Layer Perceptron (MLP) or Principal Component Analysis (PCA).

\subsection{Joint Behavioral Modeling}
Existing semantic-enhanced CDSR methods directly employ semantic embeddings $\mathbf{X}^{A}$ and $\mathbf{X}^{B}$ as item representations for cross-domain behavioral modeling. However, these representations are often isolated across domains and cannot reliably establish cross-domain behavioral relations. To alleviate such isolation, we instead learn behavioral representations by jointly optimizing randomly initialized item embeddings $\mathbf{E}^{A}$ and $\mathbf{E}^{B}$ together with a shared sequential encoder $f_{\tilde{\theta}}(\cdot)$.

To this end, we introduce two complementary objectives. Local Semantic Association (LSA) is motivated by the observation that semantically similar items are more likely to exhibit consistent behavioral transitions, and therefore uses local semantic relations to regularize intra-domain learning. Global Persona Guidance (GPG) instead leverages transferable user personas as high-level behavioral abstractions, encouraging representations from different domains to evolve under a shared rationale. Together, the two objectives learn behavioral representations with enriched local patterns and improved cross-domain transferability.

\noindent \textbf{Local Semantic Association.} 
Rather than globally transferring semantic knowledge across domains, LSA exploits local semantic relations to regularize intra-domain representation learning. Taking Domain~A as an example, we first construct the local semantic neighborhood of each item $i$ from the frozen semantic embeddings $\mathbf{X}^{A}$:
\begin{equation}\label{eq:local_neighbor}
\mathcal{N}^{A}_{(i)}=\operatorname{TopK}\!\left(\{c_{ij}^{A}\}_{j\in\mathcal{I}^{A},\,j\neq i}\right),\,\,c_{ij}^{A}=\cos(\mathbf{x}_i^A,\mathbf{x}_j^A),
\end{equation}
where $\mathcal N_{(i)}^{A}$ denotes the Top-$K$ semantic neighbors of item $i$. We then measure the corresponding local relations in the learned behavioral embeddings as $\hat{c}_{ij}^{A}= \cos(\mathbf{e}_i^A,\mathbf{e}_j^A)$, and minimize the discrepancy between semantic similarity and behavioral similarity within each local neighborhood. Thus:
\begin{equation}\label{eq:loss_lsa} \mathcal{L}_{lsa}^{A}=\frac1{|\mathcal B^{A}|}\sum_{i\in\mathcal B^{A}}\frac1{|\mathcal N^{A}_{(i)}|}\sum_{j\in\mathcal N^{A}_{(i)}}(\hat c_{ij}^{A}-c_{ij}^{A})^2,
\end{equation}
where $\mathcal B^{A}$ denotes the unique items of domain A in the current batch. The same procedure is applied to domain~B, and the final loss is $\mathcal{L}_{lsa}=\frac{1}{2}(\mathcal{L}_{lsa}^{A}+\mathcal{L}_{lsa}^{B})$.

\noindent \textbf{Global Persona Guidance.} 
To inject the inferred persona into sequential behavior modeling, we first encode the mixed-domain interaction sequence $\tilde{s}_u$ using the shared sequential encoder $f_{\tilde{\theta}}(\cdot)$, yielding hidden representations $\{\tilde{\mathbf{h}}_u^{(t)}\}_{t=1}^{|\tilde{s}_u|}$. Since different interactions contribute unequally to the persona, we employ a persona-aware attention mechanism to estimate the importance of each hidden representation according to its similarity with the persona anchor $\mathbf{p}_u$:
\begin{equation}\label{eq:persona_weight}
w_u^{(t)}=\frac{\exp(\cos(\tilde{\mathbf{h}}_u^{(t)},\mathbf{p}_u))}{\sum_{k=1}^{T}\exp(\cos(\tilde{\mathbf{h}}^{(k)}_u,\mathbf{p}_u))}.
\end{equation}
We then optimize the behavioral representations toward their corresponding persona through weighted contrastive guidance. Specifically, each hidden representation is encouraged to align with its own persona while being distinguished from other users' personas. The objective is formulated as:
\begin{equation}\label{eq:loss_gpg}
    \mathcal{L}_{gpg}=\frac{-1}{|\mathcal{B}|}\sum_{u\in\mathcal{B}}\sum_{t=1}^{|\tilde{s}_u|}w_u^{(t)}\left(\log\frac{\exp(\cos(\tilde{\mathbf{h}}^{(t)}_u,\mathbf{p}_u)/\tau)}{\sum\limits_{v\in\mathcal{B}}\exp(\cos(\tilde{\mathbf{h}}^{(t)}_u,\mathbf{p}_v)/\tau)}\right),
\end{equation}
where $\mathcal{B}$ denotes the unique users in the current batch, and $\tau$ is the temperature coefficient. Consequently, behavior representations from different domains are gradually optimized toward the same persona anchor, allowing transferable cross-domain behavioral relations to emerge naturally.

\subsection{Domain Preference Adaptation}
Through joint behavioral modeling, the mixed-domain interaction sequence is encoded as $\tilde{\mathbf h}_u=f_{\tilde{\theta}}(\tilde{\mathbf E}[\tilde{s}_u])$, where $\tilde{\mathbf E}$ denotes the mixed behavioral embedding table constructed from $\mathbf E^A$ and $\mathbf E^B$. The joint scores are computed as $\tilde{\mathbf r}_u^{A}=\tilde{\mathbf h}_u(\mathbf E^{A})^\top$ and $\tilde{\mathbf r}_u^{B}=\tilde{\mathbf h}_u(\mathbf E^{B})^\top$. While the joint modeling captures shared behaviors across domains, they inevitably sacrifice domain-aware preferences. We therefore leverage the frozen semantic embeddings, which preserve domain-specific semantics, to recover domain-aware preferences and refine the joint predictions.

\noindent \textbf{Domain Preference Modeling.} 
Taking Domain~A as an example, the domain-aware scores are computed as $\mathbf{r}_u^{A}=\mathbf{h}_u^A(\mathbf{X}^{A})^\top$, where $\mathbf{h}_u^{A}=f_{\theta}^{A}(\mathbf{X}^{A}[s_u^{A}])$ is the representation learned by an independent sequential encoder $f_{\theta}^{A}(\cdot)$.

\noindent \textbf{Adaptive Score Refinement.} We further introduce an adaptive gating network $g_{\theta}^{A}(\cdot)$ to determine the contribution $\alpha_u^A$ of domain-aware preferences:
\begin{equation}\label{eq:gate_dpa}
\alpha_u^A=g_\theta^A(\tilde{\mathbf h}_u\mathbin{\|}\mathbf h_u^{A}),
\end{equation}
where $g_{\theta}^{A}(\cdot)$ is implemented as a two-layer MLP with a Sigmoid activation and $\|$ denotes vector concatenation. The final refined scores are then obtained as $\hat{\mathbf y}_u^{A}=\tilde{\mathbf r}_u^{A}+\alpha_u^A\mathbf r_u^{A}$. The refinement process for domain~B is performed analogously.

\begin{table*}[htbp]
\centering
\resizebox{\textwidth}{!}{
\begin{tabular}{llcccccccccccc}
\toprule
\toprule
\multirow{2}{*}{\textbf{Datasets}} &  \multirow{2}{*}{\textbf{Metrics}} & \multicolumn{3}{c}{\textbf{SDSR Methods}} & \multicolumn{3}{c}{\textbf{CDSR Methods}} & \multicolumn{4}{c}{\textbf{Semantic-enhanced CDSR Methods}} & \multicolumn{1}{c}{\textbf{Ours}} & \multirow{2}{*}{\textbf{Imp.}} \\
\cmidrule(lr){3-5}\cmidrule(lr){6-8}\cmidrule(lr){9-12}\cmidrule(lr){13-13}& & GRU4Rec & Bert4Rec & SASRec & CD-SASRec & C$^2$DSR & Tri-CDR & UniSRec & CTT & LLM4CDSR & SemaCDR & PerCDR \\
\midrule
\multirow{2}{*}{\textbf{CD}} & H@10 & 0.2582 & 0.2666 & 0.3164 & 0.3142 & 0.3302 & 0.3427 & 0.3785 & 0.3945 & 0.4181 & \underline{0.4479} & \textbf{0.4861$^*$} & 8.53\% \\
& N@10 & 0.1411 & 0.1484 & 0.1979 & 0.1931 & 0.1897 & 0.2016 & 0.2271 & 0.2296 & 0.2595 & \underline{0.2770} & \textbf{0.2971$^*$} & 7.26\% \\
\midrule
\multirow{2}{*}{\textbf{Movie}} & H@10 & 0.3615 & 0.3132 & 0.3494 & 0.3748 & 0.3858 & 0.3918 & 0.4481 & 0.4524 & 0.5056 & \underline{0.5074} & \textbf{0.5519$^*$} & 8.77\% \\
& N@10 & 0.2188 & 0.1875 & 0.2092 & 0.2340 & 0.2300 & 0.2354 & 0.2732 & 0.2791 & 0.3144 & \underline{0.3182} & \textbf{0.3672$^*$} & 15.40\% \\
\midrule
\midrule
\multirow{2}{*}{\textbf{Food}} & H@10 & 0.4130 & 0.4170 & 0.4344 & 0.4328 & 0.4446 & 0.4528 & 0.4632 & 0.4787 & \underline{0.4902} & 0.4762 & \textbf{0.5247$^*$} & 7.04\% \\
& N@10 & 0.2539 & 0.2681 & 0.2796 & 0.2821 & 0.2761 & 0.2820 & 0.2914 & 0.3029 & \underline{0.3201} & 0.3170 & \textbf{0.3402$^*$} & 6.28\% \\
\midrule
\multirow{2}{*}{\textbf{Kitchen}} & H@10 & 0.3584 & 0.3661 & 0.3727 & 0.3802 & 0.3959 & 0.4062 & 0.4116 & 0.4108 & \underline{0.4268} & 0.4172 & \textbf{0.4355$^*$} & 2.04\% \\
& N@10 & 0.2100 & 0.2158 & 0.2209 & 0.2277 & 0.2376 & 0.2349 & 0.2319 & 0.2368 & \underline{0.2582} & 0.2463 & \textbf{0.2693$^*$} & 4.30\% \\
\midrule
\midrule
\multirow{2}{*}{\textbf{Cloth}} & H@10 & 0.3815 & 0.3856 & 0.4028 & 0.4250 & 0.4353 & 0.4388 & 0.4852 & 0.4982 & \underline{0.5243} & 0.5189 & \textbf{0.5552$^*$} & 5.89\% \\
& N@10 & 0.2246 & 0.2247 & 0.2649 & 0.2720 & 0.2685 & 0.2695 & 0.2874 & 0.2999 & \underline{0.3259} & 0.3183 & \textbf{0.3724$^*$} & 14.27\% \\
\midrule
\multirow{2}{*}{\textbf{Sport}} & H@10 & 0.3989 & 0.3845 & 0.4113 & 0.4272 & 0.4361 & 0.4427 & 0.4957 & 0.4930 & \underline{0.5215} & 0.5083 & \textbf{0.5567$^*$} & 6.75\% \\
& N@10 & 0.2271 & 0.2172 & 0.2591 & 0.2723 & 0.2582 & 0.2639 & 0.2963 & 0.2898 & \underline{0.3239} & 0.3130 & \textbf{0.3583$^*$} & 10.62\% \\
\bottomrule
\bottomrule
\end{tabular}
}
\caption{Overall performance comparison. The best and second-best results are bold and underlined, respectively. “Imp.” denotes the relative improvement over the strongest baseline, and “*” indicates statistical significance at the 0.05 level (paired t-test).}
\label{tab:main_results}
\end{table*}

\subsection{Model Optimization}
\noindent \textbf{Recommendation loss function.} Building upon the final prediction scores $\hat{\mathbf y}_u^{A}$ and $\hat{\mathbf y}_u^{B}$, we optimize the recommendation objective using the binary cross-entropy (BCE) loss:
\begin{equation}\label{eq:loss_rec}
    \mathcal{L}_{rec}=-\sum_{(u,i)}\left[y_{u,i}\log\sigma(\hat y_{u,i})+(1-y_{u,i})\log\!\left(1-\sigma(\hat y_{u,i})\right)\right],
\end{equation}
where $y_{u,i}\in\{0,1\}$ indicates whether item $i$ is the ground-truth next item for user $u$, $\hat{y}_{u,i}$ is the final predicted score, and $\sigma(\cdot)$ denotes the Sigmoid function.

\noindent \textbf{Overall optimization objective.} The proposed framework is optimized by jointly minimizing the recommendation loss together with two cross-domain auxiliary objectives:
\begin{equation}\label{eq:loss_all}
\mathcal{L}=\mathcal{L}_{rec}+\eta\mathcal{L}_{lsa}+\lambda\mathcal{L}_{gpg},
\end{equation}
where $\eta$ and $\lambda$ are balancing hyperparameters.

\noindent \textbf{\textit{See the Appendix for detailed algorithms and discussions.}}

\section{Experiments}
\subsection{Experimental Settings}
\textbf{Datasets.} We conduct experiments on three widely used cross-domain datasets constructed from Amazon\footnote{https://jmcauley.ucsd.edu/data/amazon/}, \textit{i.e.}, CD–Movie, Food–Kitchen, and Cloth–Sport. Following previous works~\cite{ma2024triple, xu2025heterogeneous}, we retain only users who appear in both domains of each dataset pair and filter out users and items with fewer than five interactions.\\
\textbf{Evaluation Protocols.} We follow the widely adopted leave-one-out evaluation and evaluate top-N performance using Hit Ratio@N (HR@N) and Normalized Discounted Cumulative Gain@N (NDCG@N)~\cite{kang2018self}. We report the results for N=10. Following common practice~\cite{liu2025bridge}, we randomly sample 99 negative items for each test instance to balance evaluation efficiency and reliability.\\
\textbf{Compared Baselines.} We compare PerCDR with three categories of representative methods: (I) \textit{Single-Domain Sequential Recommendation (SDSR) Methods}, which model user behaviors independently within each domain, including GRU4Rec~\cite{jannach2017recurrent}, BERT4Rec~\cite{sun2019bert4rec}, and SASRec~\cite{kang2018self}. (II) \textit{CDSR Methods}, which explicitly model transferable user behaviors across domains, including CD-SASRec~\cite{alharbi2021cross}, C$^2$DSR~\cite{cao2022contrastive}, and Tri-CDR~\cite{ma2024triple}. (III) \textit{Semantic-Enhanced CDSR Methods}, which leverage semantic information to facilitate cross-domain modeling, including UniSRec~\cite{hou2022towards}, CTT~\cite{zhou2025contrastive}, LLM4CDSR~\cite{liu2025bridge}, and SemaCDR~\cite{zhang2026semacdr}.\\
\textbf{Implementation Details.} For a fair comparison, all methods are trained under the same experimental settings. Specifically, the hidden dimension is set to 128, the batch size to 256, the maximum number of training epochs to 200 with an early stopping patience of 10, the learning rate to 0.001, the dropout rate to 0.2, and the maximum sequence length to 50. All models are optimized using the Adam optimizer. To ensure both fairness and reproducibility, we consistently employ the same open-source Qwen models\footnote{https://huggingface.co/Qwen/models} for all semantic-enhanced CDSR baselines and PerCDR. Specifically, Qwen3-Embedding-0.6B is adopted for semantic embedding generation, while Qwen3-4B is used for text reasoning. For PerCDR, the neighborhood size and temperature coefficient are fixed to $K=10$ and $\tau=0.01$, respectively. Only the balancing hyperparameters $\eta$ and $\lambda$ are tuned from $\{0.01, 0.05, 0.1, 0.5, 1, 5, 10\}$ based on the validation set.

\subsection{Overall Performance}
Table~\ref{tab:main_results} reports the overall performance on three cross-domain datasets. We summarize the following observations:\\
(I) Compared with SDSR methods, CDSR methods consistently achieve better performance by explicitly modeling transferable user behaviors across domains, demonstrating the effectiveness of cross-domain behavioral collaboration for sequential recommendation.\\
(II) Compared with classic CDSR methods, semantic-enhanced CDSR methods further improve performance by exploiting item semantics to establish richer cross-domain relations. Particularly, recent methods, \textit{e.g.}, LLM4CDSR and SemaCDR, further exploit LLM reasoning to extract richer semantic information, leading to additional gains.\\
(III) Our PerCDR consistently outperforms all semantic-enhanced CDSR methods. Existing methods globally model cross-domain behaviors from an item-centric semantic perspective. In contrast, PerCDR adopts a user-centric perspective by leveraging LLM-inferred personas to guide cross-domain behavioral modeling, while exploiting only local semantic information for intra-domain enhancement.

\subsection{In-depth Analysis}

\begin{figure}[ht]
\centering
\includegraphics[width=\columnwidth]{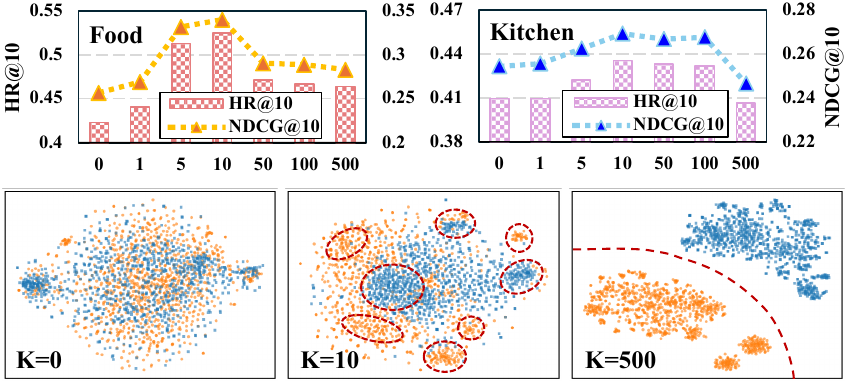} 
\caption{(Top) Performance with different semantic neighborhood sizes $K$ in LSA. (Bottom) t-SNE visualization of behavioral embeddings learned with different $K$.}
\label{pic:IA_LSA}
\end{figure}

\noindent \textbf{Necessity of Local Semantic Association (LSA).} Figure~\ref{pic:IA_LSA} investigates the effect of the semantic neighborhood size $K$. Moderate neighborhoods, \textit{e.g.}, $K=10$, consistently improve performance and produce locally structured behavioral embeddings without introducing obvious domain characteristics, indicating that LSA enhances intra-domain behavioral modeling without disrupting transferable behaviors. In contrast, an excessively large neighborhood, \textit{e.g.}, $K=500$, drives the item embeddings back to domain-specific clusters, resembling semantic isolation and causing significant performance degradation. This validates our design of leveraging local semantic relations for intra-domain modeling rather than globally bridging two domains.

\begin{figure}[t]
\centering
\includegraphics[width=\columnwidth]{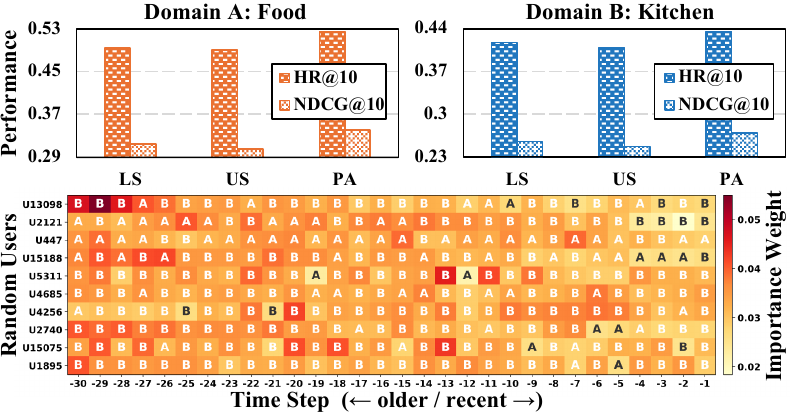} 
\caption{(Top) Performance under different persona supervision strategies in GPG. (Bottom) Heatmap of persona supervision weights over different users and time steps.}
\label{pic:IA_GPG}
\end{figure}

\noindent \textbf{Mechanism of Global Persona Guidance (GPG).} We compare three persona supervision strategies in Eq.(\ref{eq:loss_gpg}): i) LS, which supervises only the Last hidden State, \textit{i.e.}, $t=|\tilde{s}_u|$; ii) US, which Uniformly supervises all hidden States, \textit{i.e.}, $w_u^{(t)}=1/|\tilde{s}_u|$; and iii) PA, our Persona-Aware supervision. LS and US suffer from insufficient supervision and excessive supervision noise, respectively, accounting for the superior performance of PA in Figure~\ref{pic:IA_GPG}. The attention heatmap further explains its advantage: the supervision weights vary significantly across users, while highly weighted states appear at different temporal positions and span both domains. This indicates that PA adaptively emphasizes user states that are more relevant to the global persona, providing more effective guidance for transferable cross-domain behavioral modeling.

\begin{figure*}[ht]
\centering
\includegraphics[width=\textwidth]{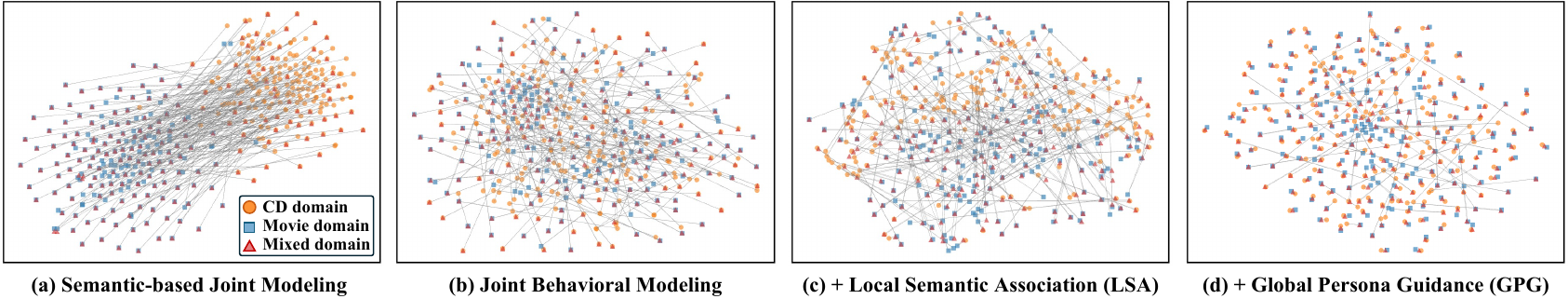} 
\caption{Evolution of cross-domain user representations under different modeling strategies on the CD-Movie dataset.}
\label{pic:IA_JBM}
\end{figure*}

\noindent \textbf{Effectiveness of Joint Behavioral Modeling (JBM).} We further visualize the relationships among the single-domain user representations and the shared representation under different modeling strategies, where the three representations of the same user are connected by gray lines, \textit{i.e.}, Figure~\ref{pic:IA_JBM}. (a) We first construct the three user representations by encoding the two single-domain sequences and the mixed-domain sequence with semantic embeddings and a shared sequential encoder. Due to semantic isolation, the two single-domain representations form distinct clusters, while the shared representation is biased toward one domain, revealing the inherent limitation of semantic-based joint modeling. (b) Replacing semantic embeddings with randomly initialized behavioral embeddings removes the explicit domain separation. However, numerous long and disordered connections indicate that representations of the same user remain weakly associated. (c) Further introducing LSA establishes local structural regularity within each domain, yet fails to effectively bridge user representations across domains. (d) Further incorporating GPG makes the three representations of the same user substantially more coherent, accompanied by a remarkable reduction in long cross-domain connections. These observations demonstrate that JBM effectively exploits LSA to enhance intra-domain behavioral modeling and GPG to establish transferable cross-domain behavioral relations.

\begin{figure}[ht]
\centering
\includegraphics[width=\columnwidth]{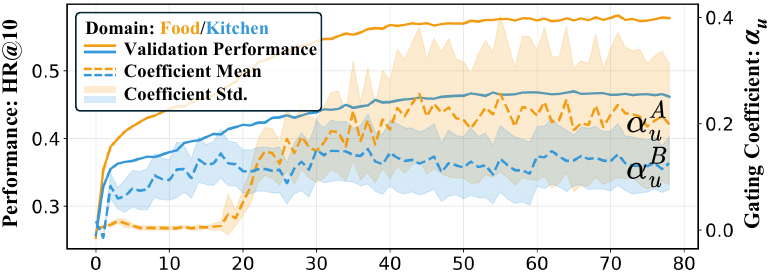} 
\caption{Dynamics of validation performance and domain-specific refinement coefficients across training epochs.}
\label{pic:IA_DPA}
\end{figure}

\noindent \textbf{Dynamics of Domain Preference Adaptation (DPA).} Figure~\ref{pic:IA_DPA} illustrates the evolution of the validation performance and the refinement coefficients $\alpha_u^A$ and $\alpha_u^B$ in Eq.(\ref{eq:gate_dpa}) during training. At the early stage, the refinement coefficients remain relatively small, indicating that the model primarily relies on the jointly learned cross-domain behavioral representations. As training proceeds, the coefficients gradually increase and stabilize, suggesting that domain-specific preference adaptation is progressively introduced after transferable behavioral relations have been established. Meanwhile, the considerable variance of the refinement coefficients across users indicates that DPA effectively balances transferable cross-domain behaviors with personalized domain-specific preferences.

\begin{figure}[ht]
\centering
\includegraphics[width=\columnwidth]{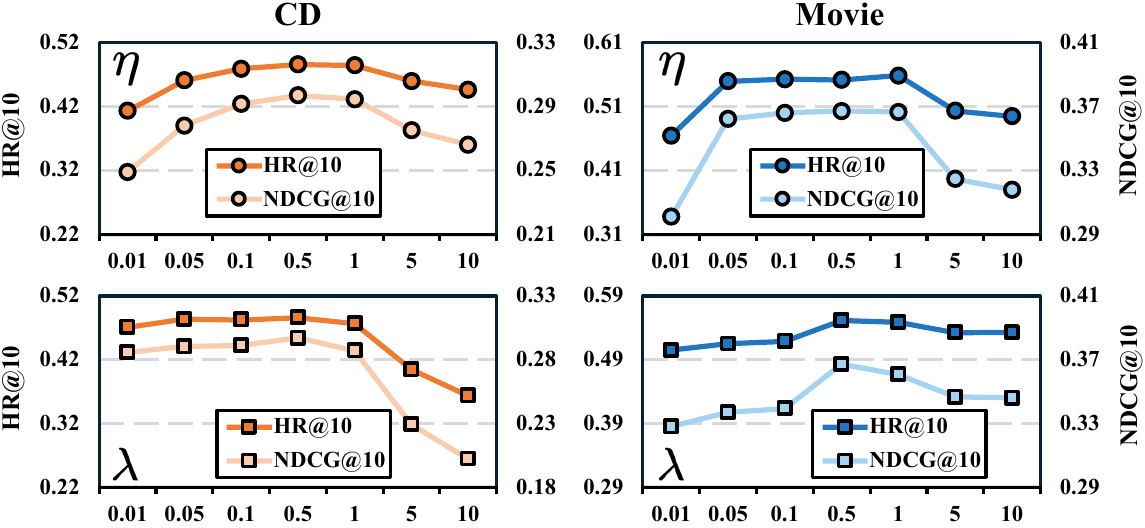} 
\caption{Effect of the loss weights $\eta$ and $\lambda$ on performance.}
\label{pic:hyperparams}
\end{figure}

\begin{table}[t]
\centering
\Large
\resizebox{\columnwidth}{!}{
\begin{tabular}{llccccc}
\toprule
\toprule
\textbf{Datasets} & \textbf{Metrics} & \textbf{PerCDR} & \textbf{w/o LSA} & \textbf{w/o GPG} & \textbf{w/o JBM} & \textbf{w/o DPA} \\
\midrule
\multirow{2}{*}{\textbf{CD}}
& H@10 & \textbf{0.4861} & 0.3822 & 0.4406 & 0.3264 & 0.3852 \\
& N@10 & \textbf{0.2971} & 0.2298 & 0.2631 & 0.1894 & 0.2160 \\
\midrule
\multirow{2}{*}{\textbf{Movie}}
& H@10 & \textbf{0.5519} & 0.4452 & 0.4734 & 0.3392 & 0.5325 \\
& N@10 & \textbf{0.3672} & 0.2841 & 0.3048 & 0.2177 & 0.3442 \\
\midrule
\midrule
\multirow{2}{*}{\textbf{Food}}
& H@10 & \textbf{0.5247} & 0.4230 & 0.4842 & 0.3168 & 0.5145 \\
& N@10 & \textbf{0.3402} & 0.2567 & 0.3058 & 0.2116 & 0.3307 \\
\midrule
\multirow{2}{*}{\textbf{Kitchen}}
& H@10 & \textbf{0.4355} & 0.4095 & 0.4077 & 0.2270 & 0.4275 \\
& N@10 & \textbf{0.2693} & 0.2545 & 0.2479 & 0.1366 & 0.2618 \\
\bottomrule
\bottomrule
\end{tabular}
}
\caption{Ablation study of PerCDR on two datasets.}
\label{tab:ablation}
\end{table}

\subsection{Ablation Study}
Building upon the in-depth analyses, we further conduct ablation studies to quantitatively verify the contribution of each component, with results reported in Table~\ref{tab:ablation}. Removing JBM leads to the largest performance degradation, demonstrating that jointly modeling transferable cross-domain behaviors is the key to the success of PerCDR. Furthermore, removing LSA or GPG also consistently degrades the performance, confirming that enhanced intra-domain behavioral modeling and transferable cross-domain behavioral relations are complementary and both indispensable for effective CDSR. Finally, removing DPA affects the final domain-specific preference refinement, resulting in a relatively smaller yet consistent performance decline.

\subsection{Hyperparameter Analysis}
We further investigate the effects of the weights $\eta$ and $\lambda$, which control the LSA loss and the GPG loss, respectively. As shown in Figure~\ref{pic:hyperparams}, the performance first improves and then declines as either weight increases. Small values provide insufficient supervision, while overly large values cause the auxiliary objectives to dominate the recommendation optimization. Overall, the model achieves competitive performance when both $\eta$ and $\lambda$ are around 0.5, and the same trend is consistently observed across other datasets.

\begin{figure}[ht]
\centering
\includegraphics[width=\columnwidth]{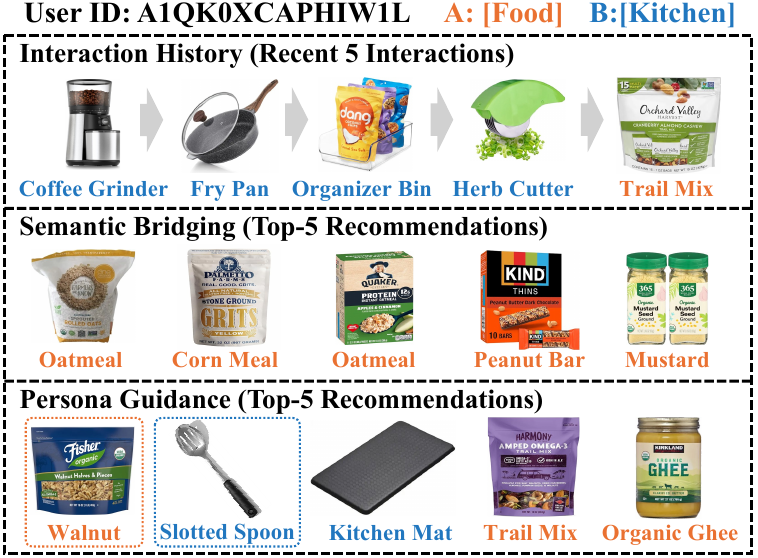} 
\caption{Case study of semantic bridging vs. PerCDR.}
\label{pic:case_study}
\end{figure}

\subsection{Case Study}
Figure~\ref{pic:case_study} presents a representative case on the Food-Kitchen dataset. Both methods perform joint modeling over the same mixed-domain interaction sequence and generate recommendations from the shared user representation, while differing in how cross-domain relations are established. Semantic-based joint modeling tends to retrieve semantically similar items concentrated in a single domain, reflecting the semantic isolation phenomenon observed in Figure~\ref{pic:motivation}(a) and Figure~\ref{pic:IA_JBM}(a). In contrast, persona guidance establishes transferable cross-domain relations through shared behavioral rationale rather than semantic proximity, enabling PerCDR to naturally recommend items from both domains and thereby alleviating the semantic-behavior inconsistency illustrated in Figure~\ref{pic:motivation}(b).

\noindent \textbf{\textit{See the Appendix for detailed settings, results, and code.}}

\section{Conclusion}
In this paper, we identify that existing CDSR methods typically adopt an item-centric perspective, relying on global semantic bridging that is inherently limited by semantic domain isolation and semantic-behavior inconsistency. We therefore shift CDSR toward a user-centric perspective and propose PerCDR, which leverages LLM-inferred personas to globally guide transferable cross-domain behavioral relations while exploiting only local semantic associations to discover intra-domain behavioral patterns. Building on this joint behavioral modeling, domain preference adaptation is further adopted for more accurate recommendations. Extensive experiments demonstrate the superiority of PerCDR.

\bibliography{aaai2027}

\newpage
\section{A~~~~Detailed Algorithm Procedure}

\begin{table}[ht]
\centering
\resizebox{1.0\columnwidth}{!}{
\begin{tabular}{cc}
\toprule
\textbf{Notation} & \textbf{Description} \\
\midrule
\multicolumn{2}{c}{\textit{Basic Sets and Sequences}} \\
\midrule
$\mathcal{U}$ 
& User set \\
$\mathcal{I}^{A},\mathcal{I}^{B}$ 
& Item sets of domain A/B \\
$\mathcal{S}^{A},\mathcal{S}^{B}$ 
& Single-domain interaction sequence sets \\
$s_u^{A},s_u^{B}$ 
& Single-domain interaction sequences of user $u$ \\
$\tilde{\mathcal{S}}$ 
& Mixed-domain interaction sequence set \\
$\tilde{s}_u$ 
& Chronologically merged sequence of user $u$ \\
\midrule
\multicolumn{2}{c}{\textit{LLM-derived Representations}} \\
\midrule
$\text{LLM}_{emb}$ & LLM embedding model \\
$\text{LLM}_{reason}$ 
&LLM reasoning model\\
$T_i$ 
& Textual attributes of item $i$ \\
$\mathbf{x}_i^{A},\mathbf{x}_i^{B}$
& Frozen semantic embeddings of item $i$ \\
$\mathbf{X}^{A},\mathbf{X}^{B}$ 
& Frozen item semantic embedding tables \\
$P_u$ 
& LLM-inferred persona description of user $u$ \\
$\mathbf{p}_u$ 
& Frozen persona anchor embedding of user $u$ \\
\midrule
\multicolumn{2}{c}{\textit{Joint Behavioral Modeling}} \\
\midrule
$\mathbf{E}^{A},\mathbf{E}^{B}$ 
& Trainable behavioral embedding tables \\
$\tilde{\mathbf{E}}$ 
& Mixed behavioral embedding table \\
$f_{\tilde{\theta}}(\cdot)$ 
& Shared sequential encoder \\
$\tilde{\mathbf{h}}_u^{(t)}$ 
& Hidden representation of user $u$ at the $t$-th position \\
$\tilde{\mathbf{h}}_u$ 
& Shared user representation \\
$\mathcal{N}_{(i)}^{A}, \mathcal{N}_{(i)}^{B}$ 
& Top-$K$ semantic neighborhoods of item $i$ \\
$c_{ij}^{A}, c_{ij}^{B}$ 
& Semantic similarity between items $i$ and $j$ \\
$\hat{c}_{ij}^{A}, \hat{c}_{ij}^{B}$ 
& Behavioral similarity between items $i$ and $j$ \\
$w_u^{(t)}$ 
& Persona-aware importance weight of the $\tilde{\mathbf{h}}_u^{(t)}$ \\
\midrule
\multicolumn{2}{c}{\textit{Domain Preference Adaptation}} \\
\midrule
$f_{\theta}^{A}(\cdot),f_{\theta}^{B}(\cdot)$ 
& Domain-specific sequential encoders \\
$\mathbf{h}_u^{A},\mathbf{h}_u^{B}$ 
& Domain-specific user representations \\
$\tilde{\mathbf{r}}_u^{A},\tilde{\mathbf{r}}_u^{B}$ 
& Joint behavioral recommendation scores \\
$\mathbf{r}_u^{A},\mathbf{r}_u^{B}$ 
& Domain-specific recommendation scores \\
$g_{\theta}^{A}(\cdot),g_{\theta}^{B}(\cdot)$ 
& Domain-specific gating networks \\
$\alpha_u^{A},\alpha_u^{B}$ 
& User-specific domain refinement coefficients \\
$\hat{\mathbf{y}}_u^{A},\hat{\mathbf{y}}_u^{B}$ 
& Final refined recommendation scores \\
\midrule
\multicolumn{2}{c}{\textit{Model Optimization}} \\
\midrule
$\mathcal{L}_{lsa}$ 
& Local Semantic Association loss \\
$\mathcal{L}_{gpg}$ 
& Global Persona Guidance loss \\
$\mathcal{L}_{rec}$ 
& Recommendation loss over the two domains \\
$\eta,\lambda$ 
& Balancing coefficients for $\mathcal{L}_{lsa}$ and $\mathcal{L}_{gpg}$ \\
$K$ 
& Number of semantic neighbors in $\mathcal{L}_{lsa}$ \\
$\tau$ 
& Temperature coefficient in $\mathcal{L}_{gpg}$ \\
$d$ 
& Dimension of embedding and representation \\
\bottomrule
\end{tabular}
}
\caption{Summary of the main notations used in PerCDR.}
\label{tab:notations}
\end{table}

\begin{algorithm}[htbp]
\caption{Offline Representation Construction}
\label{alg:offline}
\small
\begin{algorithmic}[1]

\REQUIRE
Item metadata $\{T_i\}$;
mixed-domain sequences $\tilde{\mathcal S}$;
LLM embedding and reasoning model $\text{LLM}_{emb}$, $\text{LLM}_{reason}$.

\ENSURE
Frozen semantic embedding tables
$\mathbf X^{A},\mathbf X^{B}$;
persona anchors
$\{\mathbf p_u\}$;
local semantic neighborhoods
$\{\mathcal N_{(i)}^{A}\},
\{\mathcal N_{(i)}^{B}\}$.

\FOR{each item $i$ in Domain $D\in\{A,B\}$}
    \STATE Encode the semantic embedding
    $\mathbf x_i^{D}$ by Eq.(\ref{eq:item_emb}).
\ENDFOR
\FOR{each item $i$ in Domain $D\in\{A,B\}$}
    \STATE Construct the Top-$K$ neighborhood
    $\mathcal N_{(i)}^{D}$ by Eq.(\ref{eq:local_neighbor}).
\ENDFOR

\FOR{each user $u$}
    \STATE Construct the persona prompt from $\tilde{s}_u$.
    \STATE Infer the persona description $P_u$.
    \STATE Encode the persona anchor
    $\mathbf p_u$ by Eq.(\ref{eq:user_anc}).
\ENDFOR

\STATE Freeze
$\mathbf X^{A},
\mathbf X^{B},
\{\mathbf p_u\}$.

\RETURN
$\mathbf X^{A},
\mathbf X^{B},
\{\mathbf p_u\},
\{\mathcal N_{(i)}^{A}\},
\{\mathcal N_{(i)}^{B}\}$.

\end{algorithmic}
\end{algorithm}

For clarity and reproducibility, Table~\ref{tab:notations} first summarizes the main notations used throughout the proposed framework. We then present the complete implementation procedure of PerCDR, which consists of three stages.

First, all LLM-related representations are constructed offline, including the frozen item semantic embeddings, user persona anchors, and local semantic neighborhoods. These representations remain fixed throughout training and therefore introduce no additional LLM inference during optimization, as summarized in Algorithm~\ref{alg:offline}.

Second, the model training procedure consists of three components: \textit{Joint Behavioral Modeling}, \textit{Domain Preference Adaptation}, and \textit{Model Optimization}. Specifically, the mixed-domain interaction sequence is first encoded to learn transferable cross-domain behavioral representations, followed by local semantic association and global persona guidance. The single-domain semantic branch is then employed to model domain-specific preferences, which are adaptively integrated with the shared behavioral branch for recommendation. Finally, all trainable parameters are jointly optimized using the overall objective defined in Eq.(\ref{eq:loss_all}). The complete training procedure is presented in Algorithm~\ref{alg:training}.

Finally, during inference, the learned shared behavioral branch first generates the joint recommendation scores, which are subsequently refined by the domain-specific semantic branch through the adaptive gating mechanism. The resulting refined scores are used for Top-$N$ recommendation in each domain, as summarized in Algorithm~\ref{alg:inference}.

\begin{algorithm}[h]
\caption{Training Procedure of PerCDR}
\label{alg:training}
\small
\begin{algorithmic}[1]
\REQUIRE Single-domain sequences $\mathcal S^{A},\mathcal S^{B}$; mixed-domain sequences $\tilde{\mathcal S}$; frozen semantic embeddings $\mathbf X^{A},\mathbf X^{B}, $; persona anchors $\{\mathbf p_u\}$; semantic neighborhoods $\{\mathcal N_{(i)}^{A}\}, \{\mathcal N_{(i)}^{B}\}$.

\ENSURE Behavioral embeddings $\mathbf E^{A}, \mathbf E^{B}$; shared encoder $f_{\tilde{\theta}}(\cdot)$; domain-specific encoders $f_{\theta}^{A}(\cdot), f_{\theta}^{B}(\cdot)$; gating networks $g_{\theta}^{A}(\cdot)$, $ g_{\theta}^{B}(\cdot)$.

\STATE Randomly initialize $\mathbf E^{A}, \mathbf E^{B}, f_{\tilde{\theta}}(\cdot), f_{\theta}^{A}(\cdot), f_{\theta}^{B}(\cdot), g_{\theta}^{A}(\cdot), g_{\theta}^{B}(\cdot)$.

\FOR{each epoch}
    \FOR{each batch}
    \STATE \texttt{<-- Joint Behavioral Modeling -->}
        \STATE Encode mixed-domain sequences and obtain shared user representations $\tilde{\mathbf h}_u$ based on $\tilde{\mathcal S}, \mathbf E^{A}, \mathbf E^{B}, f_{\tilde{\theta}}(\cdot)$ by Eq.(\ref{eq:seq_rec}).

        \STATE Compute the Local Semantic Association loss $\mathcal L_{lsa}$ according to Eq.(\ref{eq:loss_lsa}).
        \STATE Compute the Global Persona Guidance loss $\mathcal L_{gpg}$ according to Eq.(\ref{eq:loss_gpg}).
    \STATE \texttt{<-- Domain Preference Adaptation -->}
        \STATE Encode single-domain sequences and obtain domain-specific user representations $\mathbf h_u^{A},\mathbf h_u^{B}$ based on $\mathcal S^{A},\mathcal S^{B}$, $\mathbf X^{A},\mathbf X^{B}, f_{\theta}^{A}(\cdot), f_{\theta}^{B}(\cdot)$ by Eq.(\ref{eq:seq_rec}).

        \STATE Compute shared behavioral scores $\tilde{\mathbf r}_u^{A}, \tilde{\mathbf r}_u^{B}$.

        \STATE Compute domain-specific semantic scores $\mathbf r_u^{A}, \mathbf r_u^{B}$.

        \STATE Compute refinement coefficients
        $\alpha_u^{A}, \alpha_u^{B}$ by Eq.(\ref{eq:gate_dpa}).

        \STATE Refine recommendation scores
        $\hat{\mathbf y}_u^{A},\hat{\mathbf y}_u^{B}$.
    \STATE \texttt{<-- Model Optimization -->}
        \STATE Compute recommendation loss $\mathcal L_{\mathrm{rec}}$ by Eq.(\ref{eq:loss_rec}).

        \STATE Compute the overall objective $\mathcal L$ by Eq.(\ref{eq:loss_all}).

        \STATE Update all trainable parameters using Adam.
    \ENDFOR
\ENDFOR

\RETURN Trained PerCDR.
\end{algorithmic}
\end{algorithm}

\begin{algorithm}[htbp]
\caption{Inference Procedure of PerCDR}
\label{alg:inference}
\small
\begin{algorithmic}[1]
\REQUIRE
Mixed-domain interaction sequence
$\tilde{s}_u$;
single-domain sequences
$s_u^{A},s_u^{B}$;
trained PerCDR.

\ENSURE
Top-$N$ recommendations for Domain A and Domain B.

\STATE Encode the mixed-domain sequence and obtain
the shared representation
$\tilde{\mathbf h}_u=f_{\tilde{\theta}}(\mathbf{\tilde{E}}(\tilde{s}_u))$.

\STATE Compute shared behavioral scores\\
$\tilde{\mathbf r}_u^{A}=\tilde{\mathbf h}_u(\mathbf{E}^{A})^{\top}$, $
\tilde{\mathbf r}_u^{B}=\tilde{\mathbf h}_u(\mathbf{E}^{B})^{\top}$.

\STATE Encode the single-domain sequences and obtain the domain-specific representations \\$\mathbf h_u^{A}=f_{\theta}^A(\mathbf{X}^A(s_u^A))$, $\mathbf h_u^{B}=f_{\theta}^B(\mathbf{X}^B(s_u^B))$.

\STATE Compute domain-specific semantic scores \\$\mathbf r_u^{A}=\mathbf h_u^A(\mathbf{X}^{A})^{\top}$, $
\mathbf r_u^{B}=\mathbf h_u^B(\mathbf{X}^{B})^{\top}$.

\STATE Compute refinement coefficients $\alpha_u^{A}, \alpha_u^{B}$
according to Eq.(\ref{eq:gate_dpa}).

\STATE Obtain the final recommendation scores
\\$\hat{\mathbf y}_u^{A}=\tilde{\mathbf r}_u^{A}+\alpha_u^A\mathbf r_u^{A}$, $\hat{\mathbf y}_u^{B}=\tilde{\mathbf r}_u^{B}+\alpha_u^B\mathbf r_u^{B}$.

\STATE Return the Top-$N$ items in each domain.
\end{algorithmic}
\end{algorithm}

\section{B~~~~Further Discussions}
\vspace{0.6em}
\noindent \textbf{\textit{B.1 Why does Semantic Bridging fall short for Cross-domain Sequential Recommendation?}}
\vspace{0.6em}

Cross-domain sequential recommendation (CDSR) fundamentally aims to \textbf{transfer user behavioral patterns across domains}. Early CDSR methods achieve cross-domain knowledge transfer by directly exploiting users’ interactions from multiple domains, where cross-domain behavioral co-occurrences naturally establish the associations between items. Such behavior-driven transfer has long been regarded as the primary mechanism for CDSR.

Recent advances further introduce semantic information with the objective of discovering additional cross-domain behavioral relations beyond observed interactions. Rather than relying solely on interaction signals, these methods expect cross-domain behaviors to emerge from globally shared semantic knowledge. Representative methods following this philosophy are summarized in Table~\ref{tab:SB_sum}, which we collectively term the \texttt{semantic bridging} paradigm.

As summarized in Table~\ref{tab:SB_sum}, despite their different implementations, these methods all rely on the same underlying assumption: \textbf{transferable cross-domain behavioral relations can be established through global semantic bridging from an \emph{item-centric} perspective}.

However, we argue that this assumption is inherently limited for two fundamental reasons:\\
\textbf{I. Semantic Domain Isolation.} Semantic representations are naturally domain-isolated due to domain-specific characteristics. As illustrated in Figure~\ref{pic:motivation}(a), Kitchen items are mainly described by functionalities, \textit{e.g.}, cooking or storage, whereas Food items are characterized by flavors, ingredients, and nutrition. Consequently, their semantic embeddings naturally form isolated semantic spaces, requiring existing methods to explicitly align, adapt, or fuse heterogeneous semantic representations before behavioral transfer, \textit{making global semantic bridging itself a challenging optimization problem}.

\noindent \textbf{II. Semantic-Behavior Inconsistency.} Global semantics do not necessarily reveal transferable behavioral relations. As illustrated in Figure~\ref{pic:motivation}(b), users often establish cross-domain behavioral transitions between semantically unrelated items, \textit{e.g.}, measuring spoons to peanut oil, or baking tools to flour. Such transitions are driven by user behavioral intentions rather than item semantic similarity. In contrast, semantic information exhibits much stronger consistency within individual domains, \textit{e.g.}, different flavors of candies or cookware designed for similar functionalities naturally correspond to similar user behaviors. Therefore, \textit{semantic information is inherently more suitable for modeling intra-domain behavioral patterns than for establishing transferable cross-domain behavioral relations}.

These observations suggest that the limitation lies in treating global semantic bridging as the primary mechanism for cross-domain transfer. Instead, semantic information is inherently better suited to preserving domain-specific preferences globally (Reason I) and enhancing intra-domain behavioral modeling locally (Reason II).

\vspace{0.6em}
\noindent \textbf{\textit{B.2 Why Does Persona Better Guide Cross-domain Behavioral Modeling?}}
\vspace{0.6em}

As discussed above, semantic bridging falls short because \textbf{CDSR should establish cross-domain behavioral relations through users’ underlying behavioral rationale rather than semantic proximity between items across domains}. Therefore, we shift from an \emph{item-centric} to a \textbf{\emph{user-centric}} perspective, directly learning transferable behavioral patterns from mixed-domain interactions.

Unlike semantic embeddings, behavioral embeddings are randomly initialized and jointly optimized over the mixed-domain interaction sequences. Consequently, they are free from domain-specific priors and naturally avoid the domain isolation discussed in Reason~I. However, interaction sequences alone provide only user-specific behavioral observations, making it difficult to establish transferable cross-domain behavioral relations. Instead, transferable relations should be guided by behavioral rationale shared across users, motivating the need for \textbf{global behavioral priors}.

To provide such global behavioral priors, PerCDR introduces \textbf{user personas rather than conventional user profiles}. Although both are high-level user abstractions, they capture fundamentally different information, as illustrated in the following example.

\begin{table*}[t]
\centering
\small
\resizebox{1.0\textwidth}{!}{
\begin{tabular}{p{1.6cm}p{13.4cm}}
\toprule
\textbf{Method} &
\textbf{Global Semantic Bridging Mechanism} \\
\midrule
\textbf{UniSRec}
(KDD'22)
&
Learns \textbf{universal semantic item representations} through domain fusion and adaptation, expecting shared semantic representations to generalize sequential behavioral patterns across domains.
\\

\textbf{CTT}
(KDD'25)
&
Transfers \textbf{cross-domain semantic relationships} by cross-attention and alignment, assuming semantic relations can establish behavioral dependencies across domains.
\\

\textbf{LLMCDSR}
(TOIS'25)
&
Performs \textbf{collaborative-textual contrastive learning} to align semantic and behavioral representations, followed by semantic-guided pseudo interaction generation to alleviate target-domain sparsity.
\\

\textbf{LLM-EDT}
(SIGIR'26)
&
Refines \textbf{semantic-behavior alignment} by denoising pseudo interactions and pretraining on mixed-domain semantic representations, producing more reliable cross-domain behavioral transfer.
\\

\textbf{LLM4CDSR}
(SIGIR'25)
&
Aligns \textbf{mixed-domain semantic representations}, domain-specific user representations, and LLM-inferred preference representations through contrastive learning.
\\

\textbf{SemaCDR}
(WWW'26)
&
Uses LLM reasoning to infer \textbf{shared item semantic attributes}, \textit{e.g.}, genre, tone, or style) between domains and exploits them to enhance potential cross-domain interactions.
\\
\bottomrule
\end{tabular}
}
\caption{Existing semantic-enhanced CDSR methods bridge cross-domain behaviors through global semantics.}
\label{tab:SB_sum}
\end{table*}

\begin{casebox}{User Profile vs. Persona}
\textbf{User ID:} A128ZZFS41K1HC

\textbf{Dataset:} Food--Kitchen

\medskip
\textbf{Item Semantics (Titles)}
\begin{list}{}{}
\item[\textbf{[Kitchen]}] DII 52x52" Square Cotton Tablecloth, Halloween Portrait - Perfect for Halloween, Dinner Parties and Scary Movie Nights
\item[\textbf{[Kitchen]}] DII Christmas Holiday Embroidered Table Runner, 14x70", Holly Leaves
\item[\textbf{[Food]}] Premier Protein 30g Protein Shakes, Chocolate, 11 Fluid Ounces, 4 Per Pack
\item[\textbf{[Kitchen]}] DII Decorative Light Up Skulls 15" Haunted Wreath for Front Door or Indoor Wall Décor to Celebrate Halloween
\item[\textbf{[Food]}] Kuli Kuli Moringa Green Energy Shots, Raspberry, Natural Energy from Green Tea and Leafy Green Superfood
\item[\textbf{[Kitchen]}] Shrunks Inflatable Kids Bed Rails, Safety Side Bumpers for Toddlers or Adult Beds
\item[\textbf{[Kitchen]}] Koziol Sahsa S Fruit Knife, Solid Cosmos Black
\item[\textbf{[Food]}] Rhythm Superfoods Organic Kale Chips Variety Pack, Vegan and Gluten-Free
\item[\textbf{[Food]}] Birch Benders Organic Chocolate Chip Pancake and Waffle Mix, Whole Grain, Non-GMO
\item[\textbf{[Food]}] Birch Benders Six Grain Cinnamon Pancake and Waffle Mix, Organic Ingredients, Whole Grain
\end{list}

\medskip
\textbf{User Profile}

Protein shakes, Healthy drinks, Organic snacks, Superfoods, Whole-grain breakfast, Kitchen utensils, Home decoration, Halloween decoration, Family household products

\medskip
\textbf{User Persona}

A family-oriented person who enjoys creating a warm and festive home, values healthy eating habits, and prefers practical products for everyday living

\end{casebox}

As shown above, user profiles primarily summarize \textbf{what users have interacted with}, resulting in high-level summaries of historical item semantics, \textit{e.g.}, preferred categories, concepts, or attributes. Consequently, they naturally \textbf{remain semantic representations} and are incorporated into semantic-bridging methods, such as LLM4CDSR and LLM-EDT, to further enhance semantic modeling.

In contrast, personas characterize \textbf{why users interact}, capturing users' underlying behavioral motivations, such as habits, intents, and consumption tendencies, which are independent of specific items or domains. Rather than enhancing semantic representations, personas \textbf{serve as global behavioral priors} shared across users. By encouraging users sharing similar personas to produce consistent optimization signals for the shared behavioral embeddings, transferable cross-domain behavioral co-occurrence can be learned beyond individual users' observed interactions.

In summary, user profiles describe the semantics of historically interacted items and are therefore primarily used to enhance semantic-bridging methods. In contrast, user personas characterize the behavioral rationale underlying user interactions, making them more suitable for guiding transferable cross-domain behavioral modeling.

\vspace{0.6em}
\noindent \textbf{\textit{B.3 Why Separate Behavioral and Semantic Representations?}} 
\vspace{0.6em}

Behavioral representations are optimized over mixed-domain interaction sequences under persona guidance, aiming to maximize \textbf{transferability across domains}. Consequently, they primarily capture domain-invariant behavioral patterns while inevitably suppressing domain-specific characteristics. In contrast, semantic representations are derived directly from item metadata and naturally preserve \textbf{domain-specific characteristics}. As discussed in Reason~I and Reason~II, these properties make semantic information particularly suitable for maintaining domain-aware preferences and reinforcing intra-domain behavioral modeling.

PerCDR therefore explicitly separates behavioral and semantic representations. The behavioral branch \textbf{learns transferable cross-domain behavioral relations}, while the semantic branch \textbf{exploits its inherent domain specificity} to refine the shared behavioral predictions. Consequently, behavioral modeling is responsible for cross-domain transfer, whereas semantic modeling restores domain-specific preferences, allowing the two objectives to be optimized without compromising each other.

\section{C~~~~More Experimental Settings}
\subsection{Datasets}
We conduct experiments on three widely used cross-domain recommendation benchmarks constructed from the Amazon Review dataset, including \textbf{CD--Movie}, \textbf{Food--Kitchen}, and \textbf{Cloth--Sport}. These dataset pairs have been extensively adopted in previous CDSR studies~\cite{liu2025bridge, zhang2026semacdr}, where the two domains exhibit different degrees of semantic relatedness and behavioral transferability. Following previous works~\cite{ma2024triple, xu2025heterogeneous}, we retain only overlapping users who have interactions in both domains and remove users and items with fewer than five interactions. For each remaining user, interactions are chronologically ordered within each domain to construct cross-domain sequential recommendation data. The resulting statistics are summarized in Table~\ref{tab:dataset_statistics}.

\begin{table}[h]
\centering
\resizebox{1.0\columnwidth}{!}{
\begin{tabular}{lccccc}
\toprule
\textbf{Dataset} & \textbf{\#User} & \textbf{\#Item} & \textbf{\#Interaction} & \textbf{Avg. Len.} & \textbf{Sparsity} \\
\midrule
CD       & \multirow{2}{*}{4,033}  & 9,506  & 46,177  & 11.45 & 99.88\% \\
Movie    & & 20,575 & 78,419  & 19.44 & 99.91\% \\
\midrule
Food     & \multirow{2}{*}{20,027} & 8,631  & 183,426 & 9.16  & 99.89\% \\
Kitchen  &  & 49,220 & 286,855 & 14.32 & 99.97\% \\
\midrule
Cloth    & \multirow{2}{*}{22,767} & 59,068 & 267,166 & 11.73 & 99.98\% \\
Sport    &  & 18,535 & 201,551 & 8.85  & 99.95\% \\
\bottomrule
\end{tabular}
}
\caption{Statistics of three cross-domain datasets.}
\label{tab:dataset_statistics}
\end{table}

\subsection{Evaluation Protocols} 
Following previous CDSR studies~\cite{kang2018self, cao2022contrastive, liu2025bridge}, we adopt the widely used leave-one-out evaluation protocol. Specifically, for each single-domain interaction sequence, the most recent interaction is used for testing, the second most recent interaction is used for validation, and all preceding interactions are used for training. For joint cross-domain modeling, the mixed-domain interaction sequence is constructed by chronologically merging the two domain-specific training sequences, rather than simply removing the last two interactions from the merged sequence. This design prevents information leakage, as interactions reserved for validation and testing are excluded before constructing the joint sequence. Consequently, all training sequences remain strictly chronological, ensuring that future interactions are never observed during training.

We evaluate recommendation performance using Hit Ratio at cutoff $N$ (HR@$N$) and Normalized Discounted Cumulative Gain at cutoff $N$ (NDCG@$N$). HR@$N$ measures whether the ground-truth item appears in the top-$N$ ranked recommendations, reflecting the retrieval capability of a model. NDCG@$N$ further considers the ranking positions of recommended items by assigning higher importance to relevant items appearing closer to the top of the recommendation list. In this paper, we report the results with $N=\{5,10\}$.

To improve evaluation efficiency while maintaining reliable comparisons, we follow the common protocol of randomly sampling 99 negative items for each test instance. The ground-truth item is ranked together with these sampled negatives~\cite{kang2018self, cao2022contrastive}.

\subsection{Compared Baselines}
For a comprehensive evaluation, we compare PerCDR with representative and SOTA methods from three categories.

\noindent\textbf{(I) Single-Domain Sequential Recommendation (SDSR).} These methods model user behaviors independently within each domain without exploiting cross-domain information.
\begin{itemize}[leftmargin=0.9em]
\item \textbf{GRU4Rec}~\cite{jannach2017recurrent}: A recurrent neural network-based SR method that adopts Gated Recurrent Units (GRUs) to model users' sequential behaviors.
\item \textbf{BERT4Rec}~\cite{sun2019bert4rec}: A bidirectional Transformer-based SR method trained with the masked item prediction objective to capture contextual dependencies in user behavior sequences.
\item \textbf{SASRec}~\cite{kang2018self}: A self-attention-based SR method that introduces the self-attention mechanism into sequential recommendation to effectively capture long-range dependencies and model user preferences.
\end{itemize}

\noindent\textbf{(II) Cross-Domain Sequential Recommendation (CDSR).}
These methods explicitly model transferable user behaviors across domains to alleviate data sparsity.
\begin{itemize}[leftmargin=0.9em]
\item \textbf{CD-SASRec}~\cite{alharbi2021cross}: An extension of SASRec for CDSR that transfers knowledge by incorporating aggregated source-domain representations into target-domain item embeddings.
\item \textbf{C$^2$DSR}~\cite{cao2022contrastive}: A contrastive learning-based CDSR method that jointly learns single-domain and cross-domain user representations by modeling intra- and inter-domain item dependencies.
\item \textbf{Tri-CDR}~\cite{ma2024triple}: A triple-sequence CDSR framework that jointly models source-domain, target-domain, and mixed-domain behavior sequences using cross-domain attention and contrastive learning to capture comprehensive and domain-specific user preferences.
\end{itemize}

\noindent\textbf{(III) Semantic-Enhanced CDSR.}\\
These methods leverage semantic information to facilitate cross-domain recommendation.

\begin{itemize}[leftmargin=0.9em]
\item \textbf{UniSRec}~\cite{hou2022towards}: A universal recommendation framework that learns transferable semantic item representations from textual descriptions and adapts them to downstream recommendation tasks.
\item \textbf{CTT}~\cite{zhou2025contrastive}: A text-enhanced CDSR method that jointly models semantic and behavioral information through dual-view learning with cross-domain semantic attention and intra-domain contrastive learning.
\item \textbf{LLM4CDSR}~\cite{liu2025bridge}: An LLM-enhanced CDSR method that exploits LLM-derived item embeddings and hierarchical user profiles, together with trainable adapters, to facilitate cross-domain knowledge transfer.
\item \textbf{SemaCDR}~\cite{zhang2026semacdr}: An LLM-enhanced CDSR method that integrates LLM-generated domain-agnostic semantics with domain-specific content and aligns them through contrastive learning using designed prompts.
\end{itemize}

\begin{table*}[t]
\centering
\resizebox{\textwidth}{!}{
\begin{tabular}{llcccccccccccc}
\toprule
\toprule
\multirow{2}{*}{\textbf{Datasets}} &  \multirow{2}{*}{\textbf{Metrics}} & \multicolumn{3}{c}{\textbf{SDSR}} & \multicolumn{3}{c}{\textbf{CDSR}} & \multicolumn{4}{c}{\textbf{Semantic-enhanced CDSR}} & \multicolumn{1}{c}{\textbf{Ours}} & \multirow{2}{*}{\textbf{Imp.}} \\
\cmidrule(lr){3-5}\cmidrule(lr){6-8}\cmidrule(lr){9-12}\cmidrule(lr){13-13}& & GRU4Rec & Bert4Rec & SASRec & CD-SASRec & C$^2$DSR & Tri-CDR & UniSRec & CTT & LLM4CDSR & SemaCDR & PerCDR \\
\midrule
\multirow{4}{*}{\textbf{CD}} & H@5 & 0.1680 & 0.1775 & 0.2309 & 0.2233 & 0.2261 & 0.2475 & 0.2730 & 0.2922 & 0.3104 & \underline{0.3328} & \textbf{0.3622$^*$} & 8.83\% \\
& N@5 & 0.1120 & 0.1199 & 0.1705 & 0.1638 & 0.1563 & 0.1709 & 0.1906 & 0.2011 & 0.2250 & \underline{0.2400} & \textbf{0.2571$^*$} & 7.13\% \\
& H@10 & 0.2582 & 0.2666 & 0.3164 & 0.3142 & 0.3302 & 0.3427 & 0.3785 & 0.3945 & 0.4181 & \underline{0.4479} & \textbf{0.4861$^*$} & 8.53\% \\
& N@10 & 0.1411 & 0.1484 & 0.1979 & 0.1931 & 0.1897 & 0.2016 & 0.2271 & 0.2296 & 0.2595 & \underline{0.2770} & \textbf{0.2971$^*$} & 7.26\% \\
\midrule
\multirow{4}{*}{\textbf{Movie}} & H@5 & 0.2551 & 0.2238 & 0.2491 & 0.2720 & 0.2899 & 0.2929 & 0.3294 & 0.3322 & 0.3737 & \underline{0.3754} & \textbf{0.4377$^*$} & 16.60\% \\
& N@5 & 0.1844 & 0.1587 & 0.1768 & 0.2011 & 0.1990 & 0.2035 & 0.2222 & 0.2294 & \underline{0.2754} & 0.2712 & \textbf{0.3302$^*$} & 19.90\% \\
& H@10 & 0.3615 & 0.3132 & 0.3494 & 0.3748 & 0.3858 & 0.3918 & 0.4481 & 0.4524 & 0.5056 & \underline{0.5074} & \textbf{0.5519$^*$} & 8.77\% \\
& N@10 & 0.2188 & 0.1875 & 0.2092 & 0.2340 & 0.2300 & 0.2354 & 0.2732 & 0.2791 & 0.3144 & \underline{0.3182} & \textbf{0.3672$^*$} & 15.40\% \\
\midrule
\midrule
\multirow{4}{*}{\textbf{Food}} & H@5 & 0.3040 & 0.3188 & 0.3319 & 0.3311 & 0.3354 & 0.3415 & 0.3524 & 0.3667 & \underline{0.3745} & 0.3727 & \textbf{0.4050$^*$} & 8.14\% \\
& N@5 & 0.2189 & 0.2246 & 0.2465 & 0.2493 & 0.2408 & 0.2460 & 0.2566 & 0.2725 & \underline{0.2873} & 0.2836 & \textbf{0.3016$^*$} & 4.98\% \\
& H@10 & 0.4130 & 0.4170 & 0.4344 & 0.4328 & 0.4446 & 0.4528 & 0.4632 & 0.4787 & \underline{0.4902} & 0.4762 & \textbf{0.5247$^*$} & 7.04\% \\
& N@10 & 0.2539 & 0.2681 & 0.2796 & 0.2821 & 0.2761 & 0.2820 & 0.2914 & 0.3029 & \underline{0.3201} & 0.3170 & \textbf{0.3402$^*$} & 6.28\% \\
\midrule
\multirow{4}{*}{\textbf{Kitchen}} & H@5 & 0.2528 & 0.2561 & 0.2664 & 0.2756 & 0.2903 & 0.2933 & 0.2919 & 0.2877 & \underline{0.3098} & 0.2985 & \textbf{0.3247$^*$} & 4.81\% \\
& N@5 & 0.1760 & 0.1836 & 0.1867 & 0.1940 & 0.2036 & 0.1984 & 0.1994 & 0.2042 & \underline{0.2196} & 0.2081 & \textbf{0.2335$^*$} & 6.33\% \\
& H@10 & 0.3584 & 0.3661 & 0.3727 & 0.3802 & 0.3959 & 0.4062 & 0.4116 & 0.4108 & \underline{0.4268} & 0.4172 & \textbf{0.4355$^*$} & 2.04\% \\
& N@10 & 0.2100 & 0.2158 & 0.2209 & 0.2277 & 0.2376 & 0.2349 & 0.2319 & 0.2368 & \underline{0.2582} & 0.2463 & \textbf{0.2693$^*$} & 4.30\% \\
\midrule
\midrule
\multirow{4}{*}{\textbf{Cloth}} & H@5 & 0.2727 & 0.2748 & 0.3104 & 0.3244 & 0.3298 & 0.3307 & 0.3446 & 0.3564 & \underline{0.4007} & 0.3953 & \textbf{0.4375$^*$} & 9.18\% \\
& N@5 & 0.1895 & 0.1853 & 0.2351 & 0.2396 & 0.2344 & 0.2345 & 0.2421 & 0.2543 & \underline{0.2897} & 0.2821 & \textbf{0.3344$^*$} & 15.43\% \\
& H@10 & 0.3815 & 0.3856 & 0.4028 & 0.4250 & 0.4353 & 0.4388 & 0.4852 & 0.4982 & \underline{0.5243} & 0.5189 & \textbf{0.5552$^*$} & 5.89\% \\
& N@10 & 0.2246 & 0.2247 & 0.2649 & 0.2720 & 0.2685 & 0.2695 & 0.2874 & 0.2999 & \underline{0.3259} & 0.3183 & \textbf{0.3724$^*$} & 14.27\% \\
\midrule
\multirow{4}{*}{\textbf{Sport}} & H@5 & 0.2746 & 0.2618 & 0.3064 & 0.3195 & 0.3179 & 0.3251 & 0.3589 & 0.3516 & \underline{0.3909} & 0.3792 & \textbf{0.4268$^*$} & 9.18\% \\
& N@5 & 0.1871 & 0.1742 & 0.2253 & 0.2377 & 0.2200 & 0.2259 & 0.2522 & 0.2443 & \underline{0.2817} & 0.2714 & \textbf{0.3162$^*$} & 12.25\% \\
& H@10 & 0.3989 & 0.3845 & 0.4113 & 0.4272 & 0.4361 & 0.4427 & 0.4957 & 0.4930 & \underline{0.5215} & 0.5083 & \textbf{0.5567$^*$} & 6.75\% \\
& N@10 & 0.2271 & 0.2172 & 0.2591 & 0.2723 & 0.2582 & 0.2639 & 0.2963 & 0.2898 & \underline{0.3239} & 0.3130 & \textbf{0.3583$^*$} & 10.62\% \\
\bottomrule
\bottomrule
\end{tabular}
}
\caption{Overall recommendation performance on six target domains. The best and second-best results are highlighted in bold and underlined, respectively. “Imp.” denotes the relative improvement over the strongest baseline, and “$^*$” indicates statistical significance at the 0.05 level based on the paired t-test.}
\label{tab:main_results_}
\end{table*}

\subsection{Implementation Details}
All experiments are conducted on a server equipped with four NVIDIA RTX A5000 GPUs, each with 24GB of memory. Unless otherwise specified, all methods are trained under the same experimental settings for a fair comparison. Specifically, the hidden dimension is set to 128, the batch size to 256, the maximum sequence length to 50, the dropout rate to 0.2, and the learning rate to 0.001. All models are optimized using the Adam optimizer and trained for at most 200 epochs with an early stopping patience of 10.

To ensure reproducibility, all baseline methods are implemented based on their official open-source code whenever available, including GRU4Rec\footnote{\url{https://github.com/hidasib/GRU4Rec}}, BERT4Rec\footnote{\url{https://github.com/FeiSun/BERT4Rec}}, SASRec\footnote{\url{https://github.com/pmixer/SASRec.pytorch}}, CD-SASRec\footnote{\url{https://github.com/XinbeiCai/CTT/tree/master/baseline/CD-SASRec}}, C$^2$DSR\footnote{\url{https://github.com/cjx96/C2DSR}}, Tri-CDR\footnote{\url{https://github.com/hulkima/Tri-CDR}}, UniSRec\footnote{\url{https://github.com/RUCAIBox/UniSRec}}, CTT\footnote{\url{https://github.com/XinbeiCai/CTT}}, LLM4CDSR\footnote{\url{https://github.com/Applied-Machine-Learning-Lab/LLM4CDSR-pytorch}}, and SemaCDR\footnote{\url{https://github.com/huangshanqiang/SemaCDR}}. Hyperparameters of all baselines are selected according to the recommendations in their original papers or official implementations, with additional tuning on the validation set whenever necessary.

To eliminate the influence of different LLM backbones, all semantic-enhanced methods, including UniSRec, CTT, LLM4CDSR, SemaCDR, and PerCDR, consistently employ the same open-source Qwen models. Specifically, Qwen3-Embedding-0.6B\footnote{\url{https://huggingface.co/Qwen/Qwen3-Embedding-0.6B}} is used to generate semantic item embeddings, while Qwen3-4B\footnote{\url{https://huggingface.co/Qwen/Qwen3-4B}} is adopted for persona reasoning and text generation. This unified configuration ensures a fair comparison by removing performance differences caused by heterogeneous foundation models rather than the recommendation algorithms themselves.

For PerCDR, the neighborhood size and temperature coefficient are fixed to $K=10$ and $\tau=0.01$, respectively. Only the balancing coefficients $\eta$ and $\lambda$ are tuned from ${0.01,0.05,0.1,0.5,1,5,10}$ according to the validation performance.

\section{D~~~~Additional Experiment Results}

\subsection{Overall Performance}

Table~\ref{tab:main_results_} presents the complete experimental results on all six target domains, including HR@5, NDCG@5, HR@10, and NDCG@10.

The overall observations are consistent with those reported in the main paper. First, CDSR methods consistently outperform single-domain sequential recommendation methods, demonstrating the effectiveness of transferring user behavioral patterns across domains. Second, incorporating semantic information further improves recommendation accuracy, and recent LLM-enhanced methods achieve stronger performance than conventional semantic-based approaches. Finally, PerCDR consistently achieves the best performance across all six target domains and all four evaluation metrics, confirming the effectiveness and robustness of the proposed persona-guided behavioral modeling framework.

Compared with conventional CDSR methods, the strongest baselines are all LLM-enhanced approaches, \textit{i.e.}, LLM4CDSR and SemaCDR, indicating the effectiveness of leveraging LLM-derived semantic knowledge for cross-domain recommendation. Nevertheless, both methods still formulate cross-domain transfer primarily from the semantic perspective. LLM4CDSR relies on global semantic item representations and LLM-summarized user profiles to facilitate cross-domain modeling. However, semantic relations do not necessarily imply behavioral transitions, causing semantically similar but behaviorally irrelevant items to be associated during representation learning.

SemaCDR further exploits manually designed prompts to infer transferable semantic characteristics, \textit{e.g.}, style or category, shared across domains. Such semantic reasoning is effective when the two domains exhibit strong semantic correlations, as evidenced by its competitive performance on the CD--Movie dataset. However, its performance becomes less competitive on semantically heterogeneous domain pairs such as Food--Kitchen, where one domain is primarily organized by functionality while the other is characterized by flavors and ingredients. Under such scenarios, transferable semantic attributes become weaker, making semantic alignment less effective for cross-domain behavioral transfer.

In contrast, PerCDR shifts the focus from semantic alignment to persona-guided behavioral modeling. Rather than explicitly searching for transferable semantics across domains, our method exploits LLM-inferred personas to supervise the learning of transferable behavioral relations while using only local semantic neighborhoods to strengthen intra-domain behavior modeling. Consequently, PerCDR consistently achieves superior performance across all datasets, especially on semantically heterogeneous domain pairs where semantic-based transfer becomes less reliable.

\begin{figure}[t]
\centering
\includegraphics[width=\columnwidth]{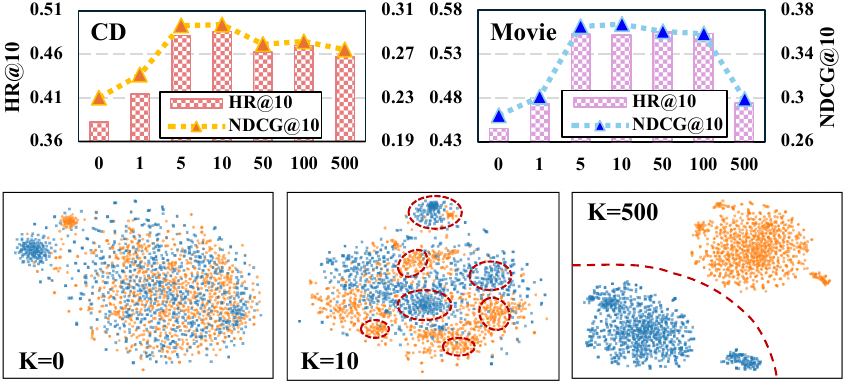} 
\caption{Additional analysis of LSA on the CD--Movie dataset. (Top) Performance with different semantic neighborhood sizes $K$ in LSA. (Bottom) t-SNE visualization of behavioral embeddings learned with different $K$.}
\label{pic:IA_LSA_cm}
\end{figure}

\subsection{In-depth Analysis}

\begin{figure}[t]
\centering
\includegraphics[width=\columnwidth]{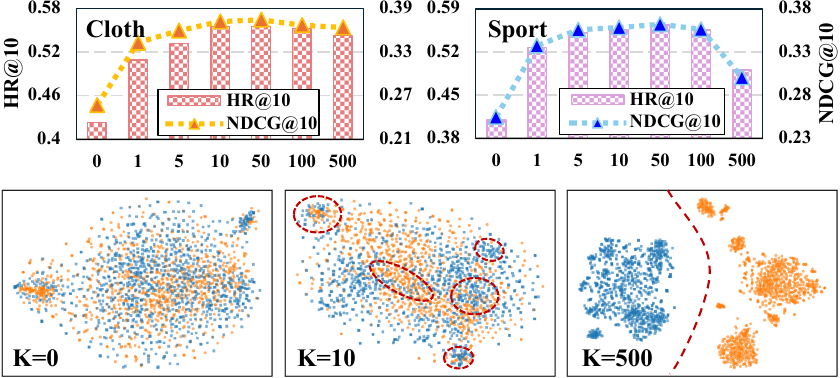} 
\caption{Additional analysis of LSA on the Cloth--Sport dataset. (Top) Performance with different semantic neighborhood sizes $K$ in LSA. (Bottom) t-SNE visualization of behavioral embeddings learned with different $K$.}
\label{pic:IA_LSA_cs}
\end{figure}

\noindent\textbf{Necessity of Local Semantic Association (LSA).} Figure~\ref{pic:IA_LSA_cm} and Figure~\ref{pic:IA_LSA_cs} present the corresponding analyses on the CD--Movie and Cloth--Sport datasets, complementing the Food--Kitchen results in the main paper. A consistent trend can be observed across all three dataset pairs: excessively small neighborhoods provide insufficient semantic guidance, whereas overly large neighborhoods gradually deteriorate the recommendation performance. This again confirms that LSA should exploit \emph{local semantic relations} rather than global semantic alignment.

Interestingly, the degradation caused by increasing $K$ varies across datasets. For domains with relatively small item spaces, such as CD and Food (<10K items), the performance drops noticeably once $K$ exceeds 10. Since each item already occupies a relatively dense semantic neighborhood, enlarging $K$ quickly expands the neighborhood from local semantic relations toward more global semantic structures, reintroducing domain-specific semantic characteristics and weakening transferable behavioral modeling. In contrast, Cloth contains nearly 60K items, resulting in a much sparser semantic space. Consequently, increasing $K$ from 10 to 500 still covers only a small fraction of the entire item space, leading to a more gradual performance degradation.

Overall, although the optimal neighborhood size is naturally influenced by the scale of the item space, a relatively moderate neighborhood, \textit{i.e.}, $K=10$, consistently achieves near-optimal performance across all datasets. These observations suggest that only a limited number of semantically related neighbors are sufficient to enhance intra-domain behavioral modeling, while introducing additional neighbors mainly injects global semantic information that is less beneficial for cross-domain behavioral transfer.

\begin{figure*}[ht]
\centering
\includegraphics[width=\textwidth]{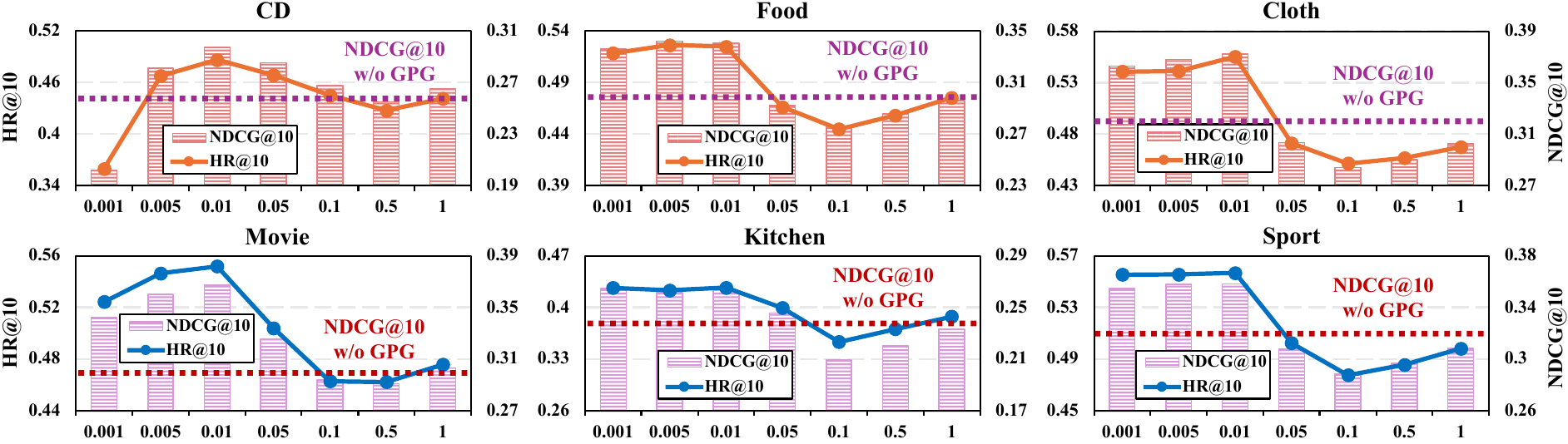} 
\caption{Sensitivity analysis of the temperature coefficient $\tau$ in GPG across all six target domains. The curves and bars represent HR@10 and NDCG@10, respectively, while the dashed lines denote the performance without GPG.}
\label{pic:IA_GPG_tau}
\end{figure*}

\begin{figure*}[ht]
\centering
\includegraphics[width=\textwidth]{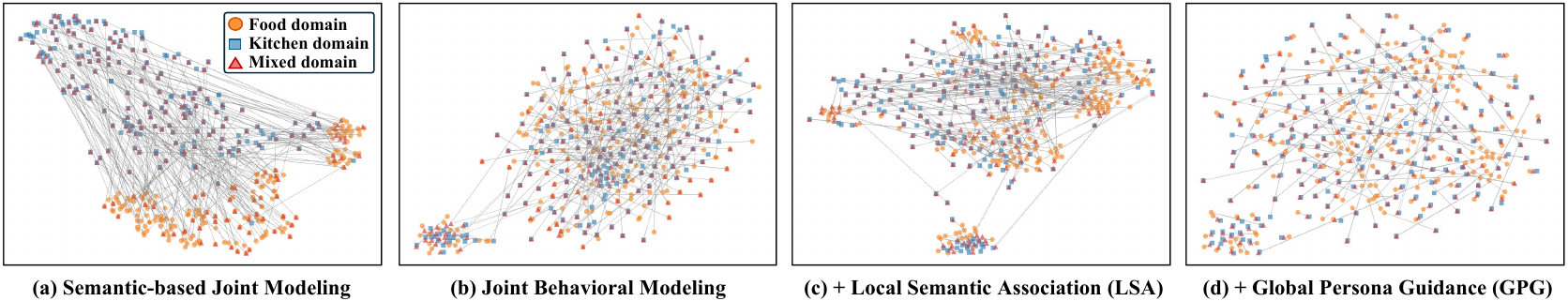} 
\caption{Evolution of cross-domain user representations under different modeling strategies on the Food-Kitchen dataset.}
\label{pic:IA_JBM_fk}
\end{figure*}

\begin{figure*}[htbp]
\centering
\includegraphics[width=\textwidth]{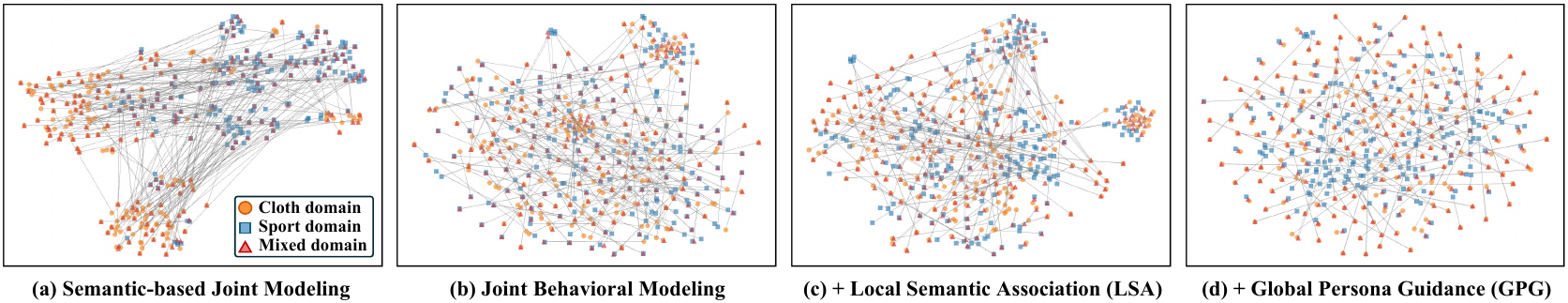} 
\caption{Evolution of cross-domain user representations under different modeling strategies on the Cloth-Sport dataset.}
\label{pic:IA_JBM_cs}
\end{figure*}

\noindent\textbf{Mechanism of Global Persona Guidance (GPG).} To further investigate the robustness of GPG, Figure~\ref{pic:IA_GPG_tau} presents the sensitivity analysis of the temperature coefficient $\tau$ in the persona-guided contrastive objective across all six domains. A highly consistent trend can be observed on all datasets. Performance first improves as $\tau$ increases from an excessively small value, reaches the optimum around $\tau=0.01$, and then gradually deteriorates as $\tau$ becomes larger.

Specifically, when $\tau$ is overly small, the contrastive objective becomes excessively sharp, causing the optimization to over-emphasize each user's own persona while suppressing informative supervision from similar personas. Consequently, the behavioral co-occurrence patterns underlying similar personas become difficult to capture, making transferable cross-domain behavioral modeling less effective. In contrast, a large temperature produces a much smoother similarity distribution, weakening the discriminative capability of the contrastive objective and gradually diminishing the effect of persona guidance. As shown in Figure~\ref{pic:IA_GPG_tau}, the recommendation performance progressively approaches that of the variant without GPG as $\tau$ increases. 

Overall, a moderately small temperature, which is also widely adopted in contrastive learning, such as $\tau=0.01$, consistently achieves near-optimal performance across all six target domains.

\noindent\textbf{Effectiveness of Joint Behavioral Modeling (JBM).} Figure~\ref{pic:IA_JBM_fk} and Figure~\ref{pic:IA_JBM_cs} present the corresponding visualizations on the Food--Kitchen and Cloth--Sport datasets, complementing the CD--Movie analysis in the main paper. Despite the substantial differences in semantic characteristics across these domain pairs, remarkably consistent representation evolution can be observed under the four modeling strategies. Specifically, semantic-based joint modeling still exhibits evident domain separation due to semantic isolation, while randomly initialized behavioral embeddings remove the explicit domain boundary but fail to establish coherent user representations. Introducing LSA further improves the local structural regularity within each domain, yet the representations of the same user remain weakly associated across domains. Finally, incorporating GPG consistently makes the three representations of each user substantially more coherent while significantly reducing long cross-domain connections.

These observations demonstrate that the effectiveness of JBM is not specific to a particular dataset. Instead, LSA consistently enhances intra-domain behavioral modeling by exploiting local semantic structures, whereas GPG establishes transferable behavioral relations through persona-guided supervision. Together, they produce coherent user representations across domain pairs with varying semantic characteristics, validating the generality of the proposed joint behavioral modeling framework.

\subsection{Ablation Study}

Table~\ref{tab:ablation_} reports the complete ablation results on all three dataset pairs, including the additional Cloth--Sport dataset. The observations are fully consistent with those reported in the main paper. Across all six domains, removing JBM consistently results in the largest performance degradation, while removing LSA, GPG, or DPA also leads to consistent performance drops, further confirming the effectiveness of every proposed component.

\begin{table}[t]
\centering
\Large
\resizebox{\columnwidth}{!}{
\begin{tabular}{llccccc}
\toprule
\toprule
\textbf{Datasets} & \textbf{Metrics} & \textbf{PerCDR} & \textbf{w/o LSA} & \textbf{w/o GPG} & \textbf{w/o JBM} & \textbf{w/o DPA} \\
\midrule
\multirow{2}{*}{\textbf{CD}}
& H@10 & \textbf{0.4861} & 0.3822 & 0.4406 & 0.3264 & 0.3852 \\
& N@10 & \textbf{0.2971} & 0.2298 & 0.2631 & 0.1894 & 0.2160 \\
\midrule
\multirow{2}{*}{\textbf{Movie}}
& H@10 & \textbf{0.5519} & 0.4452 & 0.4734 & 0.3392 & 0.5325 \\
& N@10 & \textbf{0.3672} & 0.2841 & 0.3048 & 0.2177 & 0.3442 \\
\midrule
\midrule
\multirow{2}{*}{\textbf{Food}}
& H@10 & \textbf{0.5247} & 0.4230 & 0.4842 & 0.3168 & 0.5145 \\
& N@10 & \textbf{0.3402} & 0.2567 & 0.3058 & 0.2116 & 0.3307 \\
\midrule
\multirow{2}{*}{\textbf{Kitchen}}
& H@10 & \textbf{0.4355} & 0.4095 & 0.4077 & 0.2270 & 0.4275 \\
& N@10 & \textbf{0.2693} & 0.2545 & 0.2479 & 0.1366 & 0.2618 \\
\midrule
\midrule
\multirow{2}{*}{\textbf{Cloth}}
& H@10 & \textbf{0.5552} & 0.4963 & 0.4825 & 0.3325 & 0.5518 \\
& N@10 & \textbf{0.3724} & 0.3321 & 0.3201 & 0.2242 & 0.3713 \\
\midrule
\multirow{2}{*}{\textbf{Sport}}
& H@10 & \textbf{0.5567} & 0.5148 & 0.5073 & 0.3592 & 0.5496 \\
& N@10 & \textbf{0.3583} & 0.3272 & 0.3231 & 0.2368 & 0.3541 \\
\bottomrule
\bottomrule
\end{tabular}
}
\caption{Complete ablation results on all six domains.}
\label{tab:ablation_}
\end{table}

\subsection{Hyperparameter Analysis}

\begin{figure}[t]
\centering
\includegraphics[width=\columnwidth]{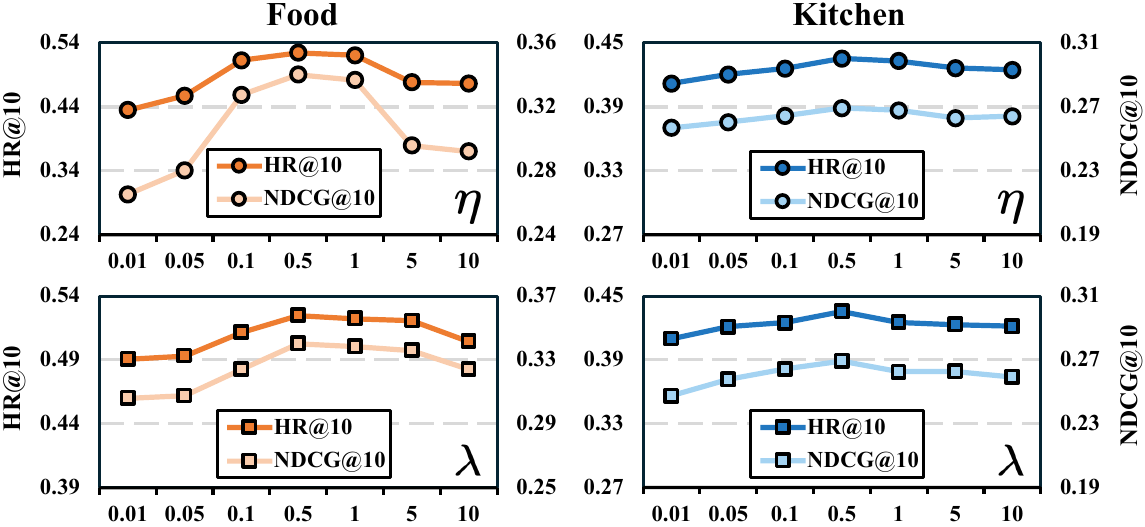} 
\caption{Effect of the loss weights $\eta$ and $\lambda$ on Food-Kitchen.}
\label{pic:hyperparams_fk}
\end{figure}

\begin{figure}[t]
\centering
\includegraphics[width=\columnwidth]{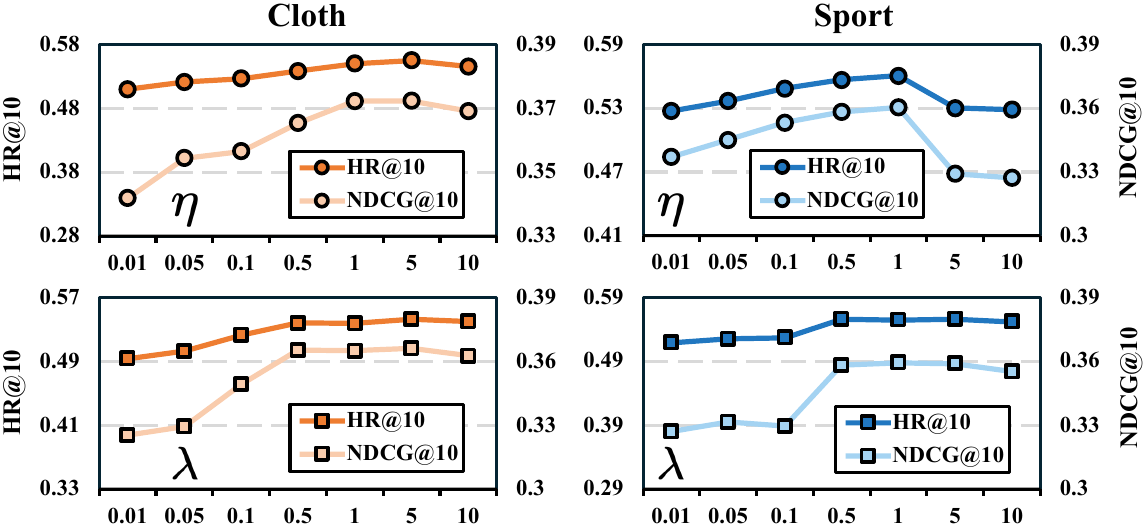} 
\caption{Effect of the loss weights $\eta$ and $\lambda$ on Cloth-Sport.}
\label{pic:hyperparams_cs}
\end{figure}

Figure~\ref{pic:hyperparams_fk} and Figure~\ref{pic:hyperparams_cs} report the hyperparameter analyses on the Food--Kitchen and Cloth--Sport datasets. Consistent with the main paper, both datasets exhibit similar performance trends. Compared with the other dataset pairs, Cloth--Sport generally favors slightly larger values of $\eta$ and $\lambda$, \textit{i.e.}, around $1.0$. Since Cloth--Sport contains substantially more users and items, stronger auxiliary supervision is beneficial for sufficiently exploiting local semantic and global persona-guided behavioral information.

\subsection{Case Study}

To further illustrate the qualitative differences between semantic-based and persona-guided cross-domain modeling, we present three representative case studies covering the CD--Movie, Food--Kitchen, and Cloth--Sport datasets. For each user, both methods perform joint modeling over exactly the same mixed-domain interaction sequence and generate recommendations from the learned shared user representation. The only difference lies in how transferable cross-domain relations are established: the semantic-bridging method relies on item semantics to associate behaviors across domains, whereas PerCDR exploits LLM-inferred personas to guide transferable behavioral modeling.

As illustrated in the three case studies, semantic bridging tends to recommend semantically related items concentrated in a single domain. This phenomenon is consistent with the semantic isolation observed in Figure~\ref{pic:IA_JBM}(a), Figure~\ref{pic:IA_JBM_fk}(a), and Figure~\ref{pic:IA_JBM_cs}(a), where the learned shared representation remains biased toward one domain. Consequently, cross-domain behavioral patterns cannot be effectively established through semantic bridging alone.

In contrast, persona guidance consistently captures cross-domain behavioral patterns beyond semantics, producing recommendations that naturally span both domains while remaining consistent with the user's historical interests. These observations further support the effectiveness and generality of the proposed persona-guided behavioral modeling framework across different cross-domain scenarios.

\begin{casebox}{Case Study 1}
\textbf{User ID:} AGBMF0IVD38U2

\textbf{Dataset:} CD-Movie

\medskip
\textbf{Recent 5 Interaction History}
\begin{list}{}{}
\item[\textbf{[Movie]}] Brian Regan: Standing Up
\item[\textbf{[Movie]}] TCM Greatest Classic Films: Legends - Esther Williams, Volume 1
\item[\textbf{[Movie]}] Ballerina
\item[\textbf{[CD]}] Dionne Warwick Collection: Her All-Time Greatest Hits
\item[\textbf{[CD]}] Tracy Chapman - Greatest Hits
\end{list}

\medskip
\textbf{Semantic Bridging (Top-5 Recommendations)}
\begin{list}{}{}
\item[\textbf{[CD]}] Playlist: The Very Best of Basia
\item[\textbf{[CD]}] '60s Gold
\item[\textbf{[CD]}] Keep Me In Your Heart For A While: The Best of Madeleine Peyroux
\item[\textbf{[CD]}] 25 Number Ones
\item[\textbf{[CD]}] Playlist: The Very Best of Kenny G
\end{list}

\medskip
\textbf{Persona Guidance (Top-5 Recommendations)}
\begin{list}{}{}
\item[\textbf{[CD]}] The Best of Gerald Levert
\item[\textbf{[CD]}] The Best of Sade
\item[\textbf{[CD]}] 20th Century Masters: The Millennium Collection - Best of Rare Earth
\item[\textbf{[Movie]}] The Survivors (VHS)
\item[\textbf{[Movie]}] David E. Talbert's Love in the Nick of Tyme
\end{list}
\end{casebox}

\begin{casebox}{Case Study 2}
\textbf{User ID:} A3BZUQXA2QHDI2

\textbf{Dataset:} Food--Kitchen

\medskip
\textbf{Recent 5 Interaction History}
\begin{list}{}{}
\item[\textbf{[Kitchen]}] Chicago Metallic Commercial II Non-Stick Mini Loaf Pans, Set of 4
\item[\textbf{[Food]}] ORCHARD VALLEY HARVEST Cranberry Almond Cashew Trail Mix, Non-GMO, No Artificial Ingredients, 1 oz (Pack of 8)
\item[\textbf{[Kitchen]}] KitchenIQ 50009 Edge Grip 2-Stage Knife Sharpener, Black, Manual
\item[\textbf{[Food]}] GoGo squeeZ YogurtZ, Banana, 3 Ounce (4 Pouches), Low Fat Yogurt, Gluten Free, Healthy Snacks, Recloseable, BPA-Free Pouches
\item[\textbf{[Kitchen]}] DII Collapsible Polyester Storage Basket with Durable Cotton Handles, Home Organizer for Office, Bedroom, Closet, Toys, Laundry (Medium - 16×10×12), Lattice Mauve
\end{list}

\medskip
\textbf{Semantic Bridging (Top-5 Recommendations)}
\begin{list}{}{}
\item[\textbf{[Kitchen]}] Tatkraft Falcon Heavy Duty Rolling Clothes Rack, Adjustable Length 3.5 to 5.7 Ft \& Height 4.5 to 6.1 Ft, Chromed Steel
\item[\textbf{[Kitchen]}] Qishi Set of 10 Portable Laundry Hanging Clothes Pins with Stainless Steel Hooks
\item[\textbf{[Kitchen]}] Uxcell a12080800ux0442 Hanging Hooks Hanger (Pack of 10)
\item[\textbf{[Kitchen]}] Eureka AS3030A Pet Upright Vacuum Cleaner, Orange
\item[\textbf{[Kitchen]}] Totally Bamboo Salt Box, "Seasoned with Love", Etched 100\% Bamboo Container with Magnetic Lid for Spices, Herbs, Seasoning \& More
\end{list}

\medskip
\textbf{Persona Guidance (Top-5 Recommendations)}
\begin{list}{}{}
\item[\textbf{[Kitchen]}] Bucky Ultra Absorbent Lightweight Lint-Free Quick-Dry Twisted Chamois Microfiber Hair Turban, Twilight
\item[\textbf{[Kitchen]}] Pre de Provence Soap Dish with a White-Washed Terracotta Finish, for Bathrooms, Kitchen Sink, Laundry, or Mud Room
\item[\textbf{[Food]}] Betty Crocker Gluten Free *The Secret Life of Pets* Fruit Flavored Snacks, 10 × 0.8 oz Pouches (Pack of 10)
\item[\textbf{[Food]}] TINY BUT MIGHTY Light Butter Microwave Popcorn, 2.5 oz (Pack of 3)
\item[\textbf{[Kitchen]}] Ultima Memory Foam Classic Pillow, Set of 2
\end{list}
\end{casebox}

\begin{casebox}{Case Study 3}
\textbf{User ID:} AS44QEHT3KSPK

\textbf{Dataset:} Cloth--Sport

\medskip
\textbf{Recent 5 Interaction History}
\begin{list}{}{}
\item[\textbf{[Sport]}] Orlimar Golf Detachable Accessory Pouch
\item[\textbf{[Sport]}] S.O.L. Survive Outdoors Longer Scout Survival Kit
\item[\textbf{[Cloth]}] Signature by Levi Strauss \& Co. Gold Label Men's Regular Fit Jeans
\item[\textbf{[Cloth]}] Fruit of the Loom Women's 3-Pack Nylon Mesh No-Show Liner Socks, Pink Heather/White (Shoe Size: 4--10)
\item[\textbf{[Sport]}] Arctix Women's Petite Muse Interchange 3-in-1 System Jacket
\end{list}

\medskip
\textbf{Semantic Bridging (Top-5 Recommendations)}
\begin{list}{}{}
\item[\textbf{[Cloth]}] Eastsport Sport Backpack for School, Hiking, Travel, Climbing, Camping, and Outdoors
\item[\textbf{[Cloth]}] Dickies Men's Original 874 Work Pants
\item[\textbf{[Cloth]}] Dickies Men's Bonded Waffle Knit Hooded Jacket
\item[\textbf{[Cloth]}] Dickies Men's Paolo Classic Chef Coat, Long Sleeve with Pearl Buttons
\item[\textbf{[Cloth]}] HDE Women's Wide Brim Floppy Packable Straw Beach Sun Hat
\end{list}

\medskip
\textbf{Persona Guidance (Top-5 Recommendations)}
\begin{list}{}{}
\item[\textbf{[Sport]}] Amerock BP53004-ORB 96-mm Center Allison Rope Pull, Oil Rubbed Bronze
\item[\textbf{[Cloth]}] Billabong Men's Classic Hybrid Shorts
\item[\textbf{[Sport]}] Phantom Aquatics Fab Series Scuba Gear Bag
\item[\textbf{[Sport]}] ColdPruf Men's Enthusiast Single-Layer Long Sleeve Crew Neck Base Layer Top
\item[\textbf{[Cloth]}] Bali Women's One Smooth U All-Over Lace Underwire Bra
\end{list}
\end{casebox}


\end{document}